\documentclass[iop]{emulateapj}
\usepackage{apjfonts}
\newcommand{\water} {H$_2$O}
\newcommand{\chtoh} {CH$_3$OH}
\newcommand{\chtcn} {CH$_3$CN}
\newcommand{\chtcnte} {CH$_3$CN (12$_K$--11$_K$)}
\newcommand{\nht} {NH$_3$}

\newcommand{\tcoto}   {$^{13}$CO(2--1)}

\newcommand{\lo}        {$L_{\odot}$}                  
\newcommand{\mo}      {$M_{\odot}$}                    
\newcommand{\kms}     {km~s$^{-1}$}                    
\newcommand{\cmth}   {cm$^{-3}$}                       
\newcommand{\cmtw}  {cm$^{-2}$}                        
\newcommand{\opac}  {cm$^2$ g$^{-1}$}                  
\newcommand{\jpbpv}     {$\rm\,Jy~beam^{-1}$ km s$^{-1}$} 

\newcommand{\Nhtw}   {$N_{\rm H_2}$}                   
\newcommand{\x}      {$\times$}                        %
\newcommand{\hii}    {H{\small II}}                    
\newcommand{\uchii}  {UC\,H{\small II}}                
\newcommand{\hchii}  {HC\,H{\small II}}                
\newcommand{\eg}     {e.\,g.,}                         
\newcommand{\mv}     {$M_{\rm vir}$}                   
\newcommand{\mgas}   {$M_{\rm gas}$}                   

\def\ca{\symbol{96}\symbol{96}}
\def\cc{\symbol{39}\symbol{39}}

\usepackage{graphicx}
\usepackage{epstopdf}
\usepackage{epsfig}
\usepackage{longtable}

\slugcomment{\sc Accepted to ApJ}

\shorttitle{SMA millimeter observations of Hot Molecular Cores}
\shortauthors{Hern\'andez-Hern\'andez et al.}

\begin{document}

\title{SMA millimeter observations of Hot Molecular Cores}

\author{Vicente Hern\'andez-Hern\'andez\altaffilmark{1}, Luis Zapata\altaffilmark{1}, 
        and Stan Kurtz\altaffilmark{1}} 
\affil{Centro de Radioastronom\'\i{}a y Astrof\'\i{}sica,
Universidad Nacional Aut\'o\-noma de M\'e\-xico,
Apdo. Postal  3--72 (Xangari), 58090 Morelia, Michoac\'an, M\'exico}
\email{V. Hern\'andez-Hern\'andez: v.hernandez@crya.unam.mx}

\and

\author{Guido Garay\altaffilmark{2}}
\affil{Departamento de Astronom\'\i{}a, Universidad de Chile,  
Camino del Observatorio 1515, Las Condes, Santiago, Chile}


\begin{abstract}

We present Submillimeter Array observations, in the 1.3\,mm continuum and the \chtcnte~
line of 17 hot molecular cores associated with young high-mass stars. The angular 
resolution of the observations ranges from 1\farcs0 to 4\farcs0. The continuum observations 
reveal large ($>$3500 AU) dusty structures with gas masses from 7 to 375 M$_\odot$, that
probably surround multiple young stars. The \chtcn~line emission is detected toward 
all the molecular cores at least up to the $K=6$-component and is mostly associated with 
the emission peaks of the dusty objects. We used the multiple $K$-components of the \chtcn~ 
and both the rotational diagram method and a 
simultaneous synthetic LTE model with the XCLASS program to estimate the temperatures and column 
densities of the cores. For all sources, we obtained reasonable fits from XCLASS by using a model 
that combines two components: an extended and warm envelope, and a compact hot core of 
molecular gas, suggesting internal heating by recently formed massive stars. 
The rotational temperatures lie in the range of 40-132\,K and 122-485\,K for
the extended and compact components, respectively. From the continuum and \chtcn~results, 
we infer fractional abundances from $10^{-9}$ to $10^{-7}$ toward the compact inner components, 
that increase with the rotational temperature. Our results agree with a chemical scenario 
in which the \chtcn~molecule is efficiently formed in the gas phase above 100-300\,K, and its abundance 
increases with temperature. 

\end{abstract}


\section{INTRODUCTION}\label{intro}

Massive stars ($M>8$\,\mo) are born inside of dense cores located
in large and massive molecular clouds \citep[e.g.,][]{1999PASP..111.1049Garay,2005IAUS..227...59C}.
These massive star-forming regions (MSFRs)
have a substantial impact on the evolution of the interstellar
medium (ISM) and make important contributions to its dynamics and
chemistry. For example, molecular outflows, jets, stellar winds and
supernovae associated with MSFRs push into their surroundings, promoting
additional star formation and mixing the ISM. 

One of the first manifestations of massive star formation is the so-called hot molecular 
core phase \citep[HMCs;][]{2000prpl.conf..299Kurtz, 2005IAUS..227...59C}. 
This phase is characterized by molecular gas condensations at relatively high temperatures
($>$100\,K) and high densities ($\sim$10$^5$--10$^8$\,\cmth), associated with a compact 
($<$ 0.1\,pc), luminous ($>$~10$^4$\,\lo), and massive ($\sim$10-1000\,\mo) molecular core.

HMCs show a forest of molecular lines, especially from organic species \citep[e.g.,][]{2005ApJS..156..127C}. 
Many of these molecules probably were formed on grain mantles during a previous cold phase, while others were 
produced by gas-phase reactions after \ca parents species\cc~were evaporated from the grains by the strong 
radiation of embedded or nearby protostars \citep[see ][]{2009ARA&A..47..427H}.

Both models and observations suggest that massive HMCs are collapsing and accreting mass 
onto a central source(s) at rates of $10^{-4}-10^{-3}$\,\mo\,yr$^{-1}$ 
\citep{2009ApJ...694...29O,2009ApJ...698.1422Z}. 
These intense mass accretion rates are high enough to prevent the development of an ionized 
region around the massive star(s) at least in the early stages \citep{1999ApJ...525..808O}. 
Thus, HMCs probably precede ultracompact \hii~regions \citep[\uchii;][]{2000prpl.conf..299Kurtz, 2001ApJ...550L..81Wilner}. 
Indeed, sub-arcsecond observations argue in favor of this scenario, particularly those showing embedded \uchii-regions, 
strong (sub)millimeter emission from dust condensations, or strong mid-IR emission from internal 
objects \citep[e.g.;][]{2010A&A...509A..50C,2011A&A...533A..73C}. 

In the above scenario a HMC 
corresponds to the most internal clump of molecular material collapsing and probably feeding 
other structures and the massive stars inside \citep{2005IAUS..227...59C,2001ApJ...550L..81Wilner}.    
However, recent sensitive high angular resolution observations suggest that the prototypical HMC, 
Orion BN/KL, may not follow this model. In this case, a close dynamical interaction of three 
young protostars produced an explosive flow and illuminated a pre-existing dense clump, 
thus creating the HMC \citep{2011A&A...529A..24Z,2011ApJ...739L..13G}. 
Also, toward G34.26+0.15 (another prototypical HMC), \citet{2007ApJ...659..447M} failed to find any 
embedded protostars within the hot core. The different nature of internally and externally heated 
HMC makes it important to distinguish between them.  

With this in mind, we present a study using Submillimeter Array\footnote{The Submillimeter Array is a 
joint project between the Smithsonian Astrophysical Observatory and the Academia 
Sinica Institute of Astronomy and Astrophysics and is funded by the Smithsonian 
Institution and the Academia Sinica.} (SMA) archival observations 
of \chtcnte~and 1.3\,mm continuum emission, toward 17 MSFRs in the HMC stage. 
\chtcn~(methyl cyanide) is frequently used as an effective thermometer and 
to estimate gas density toward HMCs \citep[e.g.,][]{2005ApJS..157..279Araya, 2001ApJ...558..194P}.
Our main goal is to use the same molecular tracer toward a relatively large group of sources 
to study the inner-most and hottest material, estimating densities, temperatures,
masses, abundances, and the spatial distribution of the dust emission and \chtcn~molecular gas.

In Section \ref{observ} we describe the archival observations presented in
this study. In Section \ref{results} we report the results and analysis of the
millimeter continuum data and the molecular line emission. In Section
\ref{comments} we comment briefly on each source, giving the 
physical characteristics from the literature and from our results. In Section \ref{discussion} 
we discuss our results, first comparing the spatial distribution of the continuum emission 
and molecular emission, and then estimating the temperatures and densities of the regions from an LTE
analysis of the \chtcnte~ spectra and through the rotation diagram method. Finally, in 
Section \ref{conclus}, we present our main conclusions.

\section{OBSERVATIONS AND DATA REDUCTION}\label{observ}

We searched the literature for MSFRs in the HMC phase, based on previous detection of 
molecular species indicating warm and dense gas such as 
\chtcn, \nht~and \chtoh. These molecules are commonly used 
to trace HMCs 
\citep[e.g.,][]{1990A&AS...83..119Churchwell, 1992A&A...253..541Churchwell, 1993A&A...276..489O, 1997A&A...321..311K, 2000A&A...354.1036K}.
We compiled a list of almost 60 objects of which most are associated with \uchii~regions, 
strong (sub)millimeter emission, molecular outflows, or maser
emission; i.e., they are young MSFRs. Then we searched in the SMA 
archive for observations that included the \chtcnte~transitions at $\sim$220.7\,GHz. 
Of the nearly 60 objects, 17 were previously observed in the compact or extended configurations and their data are public.

In Table~\ref{sources} we list the observed sources, their
coordinates, {\it lsr} velocities, distances, and luminosities, and indicate whether a \uchii~region is present.
If the 1.3\,mm continuum or \chtcnte~data were previously published, we list the paper 
in Table~\ref{obspar}. 
Distances range from 1 to 8.5\,kpc, with a mean of 5.0\,kpc. Luminosities were in most cases estimated 
from IRAS fluxes and range from 10$^4$ to some 10$^5$\,\lo. Twelve of the MSFRs (70\%) host \uchii~regions.

The HMCs were observed with the SMA \citep{2004ApJ...616L...1H} in the compact and/or extended
configuration at epochs from April 2004 to April 2010. The maximum
projected baselines of the visibility data ranged from $\sim$53 to
$\sim$174\,k$\lambda$, with different numbers of antennas at different
epochs. The SMA correlator was operated with the double-sideband receiver
covering 2\,GHz in both the lower and upper sidebands. For G23.01 a single receiver with 
4\,GHz bandwidth was used. The lower-sideband (LSB) covered the frequencies of the \chtcnte~$K$-components,
which range from 220.74726\,GHz ($K=0$) to 220.2350\,GHz ($K=11$), with uniform
spectral resolution of 0.406 MHz ($\sim$0.53\,\kms) or 0.812\,MHz ($\sim$1.1\,\kms) for different 
sources (see Table~\ref{obspar}). 
The primary beam of the SMA at 220\,GHz has FWHM $\sim$ 55\farcs

The gain, flux, and bandpass calibrators used at each epoch are listed in Table~\ref{obspar}. 
Based on the SMA monitoring of quasars, we estimate 
the uncertainty in the fluxes to be between 15\%~and 20\%.  The visibilities from each
observation were calibrated with the IDL-based MIR package (adapted for the 
SMA\footnote{The MIR cookbook by Charlie Qi at http://www.cfa.harvard.edu/$\sim$cqi/mircook.html}), 
and were then exported to MIRIAD for further processing. The 1.3\,mm continuum
emission was derived from the line-free channels of the LSB in the
visibility domain. All the line data were smoothed to a spectral resolution of 0.812 MHz or $\sim$1.1\,\kms to improve the sensitivity and provide uniform spectra. 
The synthesized beam sizes range from 1\farcs47\x0\farcs83 to 
5\farcs34\x2\farcs95. In Table~\ref{obspar} we summarize the relevant information concerning the observations.

\section{RESULTS AND ANALYSIS}\label{results}

\subsection{Millimeter continuum data}\label{millcont}

In Figures \ref{lines1} to \ref{lines3} we show the 1.3\,mm continuum emission images overlaid with 
three $K$-lines ($K=3$, 5 and 7) of \chtcnte~emission toward the 17 HMCs.
Table \ref{mmcont} shows the corresponding continuum emission parameters, derived using line-free channels from the LSB. 
Using the task {\tt imfit} in MIRIAD, we found the position of the peak, and the peak and integrated flux densities. 
The deconvolved source sizes were determined from two dimensional Gaussian fits.

Since HMCs are chemically rich, the continuum emission may be contaminated by some molecular lines, 
particularly for extremely rich sources such as I17233, G10.62, G31.41, W51e2 and W51e8. 
Although we were careful to avoid any obvious contamination during the reduction process, we consider 
the peak and integrated fluxes as upper limits.

Some HMCs show embedded or very nearby \uchii~regions and the 1.3\,mm continuum emission may have 
contributions from both ionized gas emission and from the dust. To estimate the free-free contribution at 
1.3\,mm we extrapolated the emission between 10 and 45\,GHz reported in the literature, 
assuming optically thin emission, i.e., considering $S_{\nu}\propto\nu^{-0.1}$. For G45.07, G45.47, and W51e2 
we choose $\sim$100\,GHz for extrapolation of the free-free emission due to their high turnover frequencies. 
In this way we derived the dust continuum emission for the HMCs which ranges from 0.31\,Jy for I18566 to 5.88\,Jy for I17233.
The measured fluxes and the contribution from thermal dust are presented in Table \ref{mmcont}.  

To estimate the gas mass and average column density we follow \citet{1983QJRAS..24..267H}. 
Assuming optically thin dust emission and a constant gas-to-dust ratio, the gas mass is:

\begin{equation}
  \label{eq:mgas}
M_{\rm gas}= \frac{S_{\rm \nu}D^{2} R_{\rm d}}{B_{\rm \nu}(T_{\rm d}) \kappa_{\rm \nu}}
\end{equation}

\noindent where $S_{\rm \nu}$, $D$, $R_{\rm d}$, $\kappa_{\rm \nu}$ and $B_{\rm \nu}(T_{\rm d})$ are the flux density, distance 
to the core, gas-to-dust ratio, the dust opacity per unit dust mass, and the Planck function 
at the dust temperature ($T_{\rm d}$), respectively. Note that $\kappa_{\rm \nu}$ ranges from 0.2 to 3.0 at 1.3\,mm,  
depending on its scaled value with frequency as  $\nu^{\rm \beta}$, where $\rm \beta$ is the dust emissivity 
index \citep[e.g.,][]{2000AJ....119.2711H,1995P&SS...43.1333H}. Following \citet{1994A&A...291..943O} and 
using $\rm \beta$=1.5, we obtain $\kappa_{1.3\,mm}=0.74$\,\opac, corresponding to a median grain size $a=0.1~\mu$m 
and a grain mass density $\rho_{\rm d}=3$\,g\,cm$^{-3}$. Our value of $\kappa_{\nu}$ is very similar to other estimates 
toward HMCs \citep[e.g.,][]{1999AJ....118..477H,2009ApJ...694...29O}.

In the Rayleigh-Jeans approximation, equation \ref{eq:mgas} gives

 \begin{equation}
  \label{eq:mgasfin}
  \left[ \frac{M_{\rm gas}}{\rm M_\odot} \right] = 432.0 \left[ \frac{S_{\nu}}{\rm Jy} \right]  \left[ \frac{D}{\rm kpc} \right]^2  \left[ \frac{T}{\rm K} \right ]^{-1} 
 \end{equation}

 \begin{equation}
  \label{eq:ngasfin}
  \left[ \frac{N_{\rm H_2}}{\rm cm^{-2}} \right]= \frac{2.35\times10^{16}}{\theta^2} \left[ \frac{S_{\nu}}{\rm Jy} \right]  \left[ \frac{T}{\rm K} \right]^{-1}
 \end{equation}

\noindent where $\theta$ is the source size in radians and we used the common gas-to-dust ratio of 100. 
At the high densities of HMCs, dust and gas are probably well-coupled through collisions and we can assume 
that they are in thermal equilibrium \citep{1998ApJ...497..276K}. Thus, we used the high temperatures derived from 
the \chtcn~(see Section \ref{xclass}) to obtain \mgas~ and \Nhtw~. These values, obtained from the estimated  
1.3\,mm dust emission, are presented in Table \ref{mmcont}.

We caution that the calculated values for the mass and column density are sensitive to both the dust emissivity index and 
the temperature assumed for each region. For example, decreasing $\beta$ to 1.0 (while keeping the same temperature) 
lowers the results by a factor of $\sim$3.3. Also, we note that distance uncertainties may be significant for some sources. 
For other sources (i.e., W3OH/TW, G5.89, G10.47, G10.62, I18182, G23.01, G45.07, and the W51 region) trigonometric 
parallaxes have been measured \citep[][]{2009ApJ...700..137R,2014arXiv1401.5377R}; these distances are more accurate.   

\subsection{Molecular line emission}\label{molline}

We detected many molecular lines toward the 17 HMCs, but the strength of each 
species and transition detected varies from source to source. We used the 
SPLATALOGUE\footnote{http://www.splatalogue.net/} website to identify by eye the main lines in 
the LSB spectra which are mostly dominated by species such as $^{13}$CO and C$^{18}$O, plus \chtcn.
Other species frequently detected were SO, SO$_2$, H$_2 ^{13}$CO, CS and HNCO. 

In Figures \ref{xclass1} and \ref{xclass2} we show the spectrum for each source, obtained from the integrated 
emission over the region of gas traced by the $K=3$ line, which is very similar to the extended component employed in 
the XCLASS program (see Section \ref{xclass}). 
We detected $K$-components in all seventeen sources at least up to the $K=5$ line, which traces gas at $\sim$247\,K. 
For six sources, I17233, G10.47, G10.62, G31.41, W51e2, and W51e8 we detected $K=8$ lines with $E_{\rm u}=525$\,K (See Table 4).
The $K=9$ line is blended with the \tcoto~line at $\sim$220.4\,GHz, making its detection ambiguous.

A complete line identification and chemical analysis is beyond the scope of this work.
Nevertheless, we note that there is substantial chemical differentiation in some sources, including pairs of 
objects as closely spaced as W3TW-W3OH or W51e2-W51e8. These differences have been explained 
as due to different physical conditions in each region, different chemical composition of ice-mantles on 
dust grains or different ages of HMCs \citep[e.g.,][]{2009ARA&A..47..427H}. 
  
In Table \ref{linedata} we present the results from the fit of Gaussian profiles to each \chtcn~$K$-component
detected. We used the CLASS software package\footnote{CLASS is part of the GILDAS software package developed by IRAM}
to estimate the line width ($\Delta$V), the integrated intensity ($\int T dv$), and the LSR velocity ($V_{\rm LSR}$) 
for each line. We show in Table \ref{linedata} the average $K$-component values for $\Delta$V and $V_{\rm LSR}$ toward each HMC.

\subsection{Temperature and density of the \chtcn~gas}\label{xclass}

\chtcn~is considered to be a good tracer of warm--hot and high density gas 
\citep[e.g.,][]{2005ApJS..157..279Araya}. Its symmetric-top molecular structure works as a rotor, 
emitting in multiple $K$-levels within a specific $J$ transition, all within a narrow 
bandwidth ($\sim$ 0.2\,GHz). This spectral characteristic is very useful in order to 
avoid certain systematic errors that occur when comparing lines of very different frequencies. The $K$-levels 
are radiatively decoupled, and are populated only through collisions \citep{1971ApJ...168L.107S}, 
thus the rotational temperature of \chtcn~is close to the kinetic temperature of the gas.  

If we assume local thermodynamic equilibrium (LTE) and optically thin gas, \chtcn~is an excellent tracer of 
kinetic temperature using methods such as rotation diagrams \citep[RDs;][]{1979ApJ...234L.139L,1991ApJS...76..617T},
population diagrams \citep[PDs;][]{1999ApJ...517..209G,2005ApJS..157..279Araya}, and simultaneous fitting of 
multiple lines in a spectrum \citep{2005ApJS..156..127C,1999pcim.conf..330S}. In general, no background 
radiation is considered, and in the RD and PD methods, uniform temperature and density are assumed.

Following the procedure outlined in \citet{1991ApJS...76..617T} and \citet{2005ApJS..157..279Araya} one obtains 
the linear equation ln$(N_{\rm u}/g_{\rm u})$= ln$(N_{\rm tot}/Q_{\rm rot})-E_{\rm u}/kT_{\rm rot}$, in 
which the slope is ($-1/T_{\rm rot}$) 
and ln$(N_{\rm tot}/Q_{\rm rot})$ the intercept. The left-hand side of this equation contains the column density per 
statistical weight of the molecular energy levels and represents the integrated intensity per statistical weight. 
In this way, plotting the natural logarithm of $N_{\rm u}/g_{\rm u}$ versus  $E_{\rm u}/k$ of each $K$-level of \chtcn~and 
calculating the best linear fit, $T_{\rm rot}$ and $N_{\rm tot}$ can be inferred. In thermodynamic equilibrium the 
rotational temperature closely approximates the kinematic temperature of the gas.

As a first approximation we estimated the column density and rotational temperature by means of the RD method. 
The rotation diagrams are shown in Figure \ref{RDs} and the results of the linear fits are listed in 
Table \ref{xclassrd}. The error bars come from the integrated intensity 
errors in the fitting to each $K$-transition with CLASS and are shown in Table \ref{linedata}.

In the case of mildly optically thick lines, RDs underestimate the upper level column density and overestimate the 
rotational temperatures. To first order, problems in these estimations can be overcome using the PD method 
which accounts for optical depth and the source filling factor as proposed in \citet[][]{1999ApJ...517..209G}. 
Finally, in the RDs method we assumed one region with a single temperature $T_{\rm rot}$.

A more sophisticated approach is to simultaneously fit multiple lines as 
outlined by \citet{2005ApJS..156..127C} and \citet{1999pcim.conf..330S}. 
Their XCLASS program\footnote{http://astro.uni-koeln.de/projects/schilke/XCLASS}
generates a synthetic spectrum for multiple molecular species, assuming multiple emission regions, 
all assumed to be in LTE. Also, XCLASS accounts for line blends and optical depth. 
For each emission region the program requires inputs for the source size, column density, rotation temperature, 
line width, and velocity offset from the $V_{\rm LSR}$. XCLASS uses the CDMS and 
JPL spectral line database \citep{1998JQSRT..60..883P,2005JMoSt.742..215M,2001A&A...370L..49M} for 
line identification. \citep[See][for details of the procedure]{2005ApJS..156..127C}. 

To form the synthetic spectra we model each source as two distinct emission regions: one extended and warm, 
with relatively low density, and the other compact and hot, with high density. The size of the regions is 
degenerate with temperature and 
column density, depending on the optical depth \citep[See Eq. 6 and 7 of][]{2005ApJS..156..127C}. 
To avoid this degeneracy, the size of the extended component was fixed and we varied the compact 
component size from 0.5 to 0.25 the size of the extended component. 
We fixed the offset from $V_{\rm LSR}$ as estimated directly from the observed spectrum (Table \ref{linedata}). 

We probed the parameter space with temperatures between 100\,K and 500\,K for the compact component, and 
from 50\,K to 150\,K for the extended component. For the column density we probed 10$^{14}$--10$^{18}$\,\cmtw~for 
the compact component, and  10$^{12}$--10$^{15}$\,\cmtw~for the extended component. 
The best fit of the synthetic to the observed spectrum was determined by a $\chi^2$ analysis. 
As we approached a better fit we used step sizes of 1\,K in temperature and $1\times10^{12}$\,\cmtw~in column density.
For $\Delta V$ we probed steps of $\pm$2, $\pm$1 and 0\,\kms~from a near value to the observed average (Table 4).
Most of the sources showed a better fit when we used larger line widths for the compact component than for 
the extended one. In order to estimate errors, we modeled new synthetic spectra perturbing separately temperatures 
and column densities 
until we measured an under/overestimation in 20\%~of the brightness temperature for the $K=2$ transition (20\% is the 
estimated upper limit flux uncertainty of observations), since such a line is mostly optically thin and not 
blended by other lines.

In Table \ref{xclassrd} we present the final fit values and in Figures \ref{xclass1} and \ref{xclass2} we 
show the observed and synthetic spectra, respectively. All HMCs showed reasonable fits with observations using 
the two-component model.

\subsection{Virial Masses}\label{vmass}
Virial mass, \mv, can be estimated using the line width and source size.  
Assuming a power-law density distribution with index $p=1.5$ in a spherical core, 
we use the expression $M_{\rm vir} = 0.40~d~\theta_{\rm CH_3CN}~\Delta V^{2}$, 
where $d$ is the distance, $\theta_{\rm CH_3CN}$ is the angular diameter and $\Delta V$ the line width, in\,kpc, 
arcsecond, and\,\kms, respectively \citep[See Eq. 1 of][]{2004A&A...426..941B,1988ApJ...333..821M}. This is the central mass assuming that the 
core has gravitationally 
bound motion. For $\Delta V$ we used the average line width from the observed spectrum (Column 3 in Table \ref{linedata}),
and for the source size we used the extended component from the XCLASS analysis (Table \ref{xclassrd}).

In the last column of Table \ref{xclassrd} we present the calculated \mv, which ranges from 60 to 473\,\mo, 
with a median of 209\,\mo. We note that most of 
the \mv~values are greater than \mgas. This imbalance would still hold even if we used the smaller 
source size of the compact component to calculate \mv.  
We caution that many of the HMCs probably are not in dynamical equilibrium owing to 
complicated kinematics, multiple star forming sites, and large 
rotating structures such as toroids and disks. Also, large optical depths, outflowing gas, and systematic 
velocity gradients will increase the line width and thus the virial mass.

\section{COMMENTS ON INDIVIDUAL SOURCES}\label{comments}

In this Section we comment the main properties of each source on the basis of previous observations and 
describe the results obtained in this work.

{\bf W3OH} is a well-known shell \uchii~region harboring OB stars at about 2.0\,kpc, 
rich in OH and \chtoh~maser emission associated with ionized gas and 
weak molecular lines \citep{1994A&A...281..505W,1995ApJ...449L..73Wilner}. We measure a 1.3\,mm 
flux density of $\sim$3.85\,Jy, similar to reported values at different wavelengths: 
3.5\,Jy at 3\,mm \citep[][]{1995ApJ...449L..73Wilner}, 3.4 and 3.6\,Jy at 1.4 and 2.8\,mm, 
respectively \citep[][]{2006ApJ...639..975C}. This is consistent with a large contribution 
of free-free emission and minimal dust emission. We estimated a gas mass of 
$\sim$19\,\mo~and column density of $3.3\times10^{24}$\,\cmtw. From interferometric observations 
of \chtcn(5-4) \citet{1994A&A...281..505W} estimated a rotation temperature of 90$\pm$40\,K.
From the LTE analysis using XCLASS we calculated rotation temperatures between 68 and 122\,K; 
consistent with the results of \citet{1994A&A...281..505W}. Notably, W3OH shows the lowest temperature in our survey. 
\chtcn~column densities of $3.3\times10^{15}$ and $7.5\times10^{13}$\,\cmtw~were estimated for the compact 
and extended regions. We estimated a \chtcn~abundance of $\sim$2$\times10^{-9}$.

{\bf W3TW} is a young source, resolved into three components by \citet{1999ApJ...514L..43Wyrowski}, 
associated with strong dust and molecular emission at (sub)millimeter wavelengths. 
Using BIMA observations, \citet{2006ApJ...639..975C} report continuum flux densities of 1.38 and 0.22\,Jy 
at 1.4 and 2.8\,mm, 
respectively. Our higher estimate of 2.67\,Jy at 1.3\,mm may result from our lower angular resolution.
\citet{2006ApJ...639..975C} found a protobinary system with a mass of $\sim$22\,\mo~ 
for the pair. Using an LTE model for the \chtcnte~emission, they found rotation
temperatures of 200\,K and 182\,K for sources A and C, respectively.
Our data cannot resolve these two sources. From the 1.3\,mm dust emission we estimate a gas mass 
of 12.4\,\mo~and H$_2$ column density of  $1.2\times10^{24}$\,\cmtw. 
Using XCLASS, we estimate temperatures of 108 and 367\,K for the extended and compact components.
We find a \chtcn~abundance of 3.2$\times10^{-8}$.

{\bf I16547} is a MSFR with a central source of 
$\sim$30\,\mo. It hosts a thermal radio jet, outflowing gas, 
knots of shocked gas and \water~masers \citep{2003ApJ...587..739Garay, 2009ApJ...701..974F}.
Although the 1.3\,mm continuum emission is extended toward the west it is dominated by a core of 
emission with the central source (see Figure 1).
Our continuum analysis ($S_{\rm \nu}^{\rm dust}\sim1.57$\,Jy) estimates a gas mass of $\sim$20\mo~toward 
the eastern core. The \chtcnte~analysis shows kinetic 
temperatures from 78 to 272\,K with XCLASS and $\sim$245\,K using RDs.
The molecular emission from $K=3,5,7$--lines is detected mainly toward the eastern 
region, coincident with the 1.3\,mm continuum peak. We estimated \chtcn~column densities of 
$2.1\times10^{16}$ and $8.8\times10^{13}$\,\cmtw~for the compact and extended regions, respectively, 
and a fractional abundance of $4.5\times10^{-8}$ toward the compact component.

{\bf I17233} shows maser emission in OH \citep{2005ApJS..160..220F}, 
\chtoh~(Walsh et al 1998) and \water~ \citep{2008AJ....136.1455Z}, and multiple 
outflows from several \hchii~regions \citep{2009A&A...507.1443L,2008AJ....136.1455Z}. 
Large-scale movement of \nht~gas suggests a rotating core \citep{2009ApJS..184..366Beuther}. 
However, SMA observations of \chtcnte~reported by \citet{2011A&A...530A..12L} show 
that this molecular tracer is probably influenced by molecular outflows. 
\citet{2011A&A...530A..12L} used XCLASS with a two-component model similar to ours and 
report temperatures of 200\,K and 50-70\,K for the compact and extended components, respectively.
Our higher values of 346\,K and 132\,K result from using smaller component sizes in the modeled spectrum.
We estimated a \chtcn~abundance of $\sim$2$\times10^{-7}$.

{\bf G5.89} is a shell-type \uchii~region probably ionized by an O5 star 	
offset $\sim$1\farcs0 from the \hii~region \citep[Feldt's star;][]{2003ApJ...599L..91F}. 
Also present are strong molecular outflows, maser activity, five 
(sub)millimeter dust emission sources, and little molecular line emission 
\citep{2008ApJ...680.1271H, 2004ApJ...616L..35S}. The locations of the five dusty objects are 
indicated in Figure \ref{lines1}. The molecular gas appears to form a cavity 
that encircles the ionized gas. Intriguingly, the peak position of the 1.3\,mm continuum emission, 
the \chtcnte~emission, and Feldt's star do not coincide. However, most of the continuum emission 
probably comes from the free-free process, instead of thermal dust. The $K=3$-line emission structure 
is much more extended than the continuum, while the $K=5$ and 6--lines trace hotter gas to the 
northeast of Feldt's star and the 1.3\,mm emission. The \chtcn~spectrum of 
G5.89 does not show emission in $K$-lines $>6$. \citet[][]{2009ApJ...704L...5S} originally reported the 
SMA \chtcnte~data. They found a decreasing temperature structure from 150 to 40\,K with respect to the 
position of Feldt's star. Using the same SMA data, we estimated temperatures of 165 and 40\,K for 
the compact and extended components. We find a fractional abundance of $3.6\times10^{-9}$ toward the 
compact component.

{\bf G8.68} is associated with the MSFR IRAS 18032-2137. Also, \water,
class II \chtoh, and OH maser emission, strong millimeter continuum
emission, but no centimeter continuum compact sources or free-free
emission are detected \citep{2011ApJ...726...97L}. Infall profiles traced with
HCO$^+$, HNC and $^{13}$CO at 3\,mm \citep{2006MNRAS.367..553P} and 
strong SiO indicative of shocks are detected \citep{1998A&AS..132..211H}. 
From the 1.3\,mm continuum analysis, \citet{2011ApJ...726...97L} estimated a 
mass of $\sim$21\,\mo~and an H$_2$ column density of at least $10^{24}$\,\cmtw,  
assuming dust temperatures of 100-200\,K. Using the same data as \citet{2011ApJ...726...97L}
but assuming a higher temperature of 281\,K, corresponding to the compact component, 
we estimate a mass of 14\,\mo~and a column density of $10^{23}$\,\cmtw. From the 
\chtcn~data \citet{2011ApJ...726...97L} estimated a 200\,K upper limit for the rotation
temperature and $10^{16}$\,\cmtw~for the column density. We obtain column densities 
of $4.2\times10^{15}$ and $1.9\times10^{14}$\,\cmtw~for the compact and extended 
components, respectively. We estimate a \chtcn~abundance of $\sim$4\x10$^{-8}$, 
which is consistent with the result of \citet{2011ApJ...726...97L}.

{\bf G10.47} is one of the brightest HMCs and nursery of several
OB stars. This source shows four \uchii~regions embedded in the hot gas
traced by \nht, \chtcn~and many other complex molecules 
\citep{1996A&A...307..599O,1998A&A...331..709C,1998A&AS..133...29H,1999A&A...341..882Wyrowski,2011A&A...536A..33R}. 
There is strong millimeter continuum emission toward two of the
\uchii~regions. We adopt a distance of 8.5\,kpc (Reid et al. 2014).
\citet{1996A&A...307..599O}, using 30 m plus PdBI merged observations of \chtcn(6-5), 
obtained rotation temperatures of 240 and 180\,K for separate spectra of a core and extended 
components, respectively. They obtained \chtcn~column densities of $6.0\times10^{16}$ and 
$3.6\times10^{15}$\,\cmtw, toward these regions. Using the NH$_3$(4,4) line \citet{1998A&A...331..709C} 
estimated kinetic temperatures of 250-400\,K toward the central regions. 
We estimate rotation temperatures of 408 and 82\,K, and \chtcn~column densities 
of $5.1\times10^{17}$ and $4.1\times10^{14}$\,\cmtw, for the compact and extended 
components, respectively. From the 1.3\,mm continuum emission, G10.47 shows the largest gas 
mass in our survey, $\sim$375\,\mo. We find a \chtcn~abundance of 7.6$\times10^{-8}$.

{\bf G10.62} is a well-studied MSFR and associated
with a \uchii~region, \water~ and OH maser emission, and multiple molecular lines, 
including \chtcn~\citep{1987ApJ...318..712Keto,1988ApJ...324..920Keto,
2005ApJ...630..987S,2005ApJ...624L..49S,2005ApJS..160..220F,2011ApJ...729..100L}. 
From the thermal dust emission and adopting a distance of 5\,kpc (Reid et al. 2014), 
we derived a mass of 116\,\mo~and $N_{\rm H_2}=6.3\times10^{23}$\,\cmtw. 
\citet{2009ApJ...703.1308K} and \citet{2011A&A...525A.151B} derived 136\,\mo~and 
82\,\mo~also using the 1.3\,mm continuum with distances of 6 and 3.4\,kpc, respectively. 
Our rotation temperatures estimated from the XCLASS program are 415 and 95\,K for the compact and 
extended components, respectively; column densities were 6.7$\times10^{15}$ and 3.0$\times10^{14}$\,\cmtw.
\citet{2011A&A...525A.151B} obtained a rotation temperature of 87\,K and column density of 
2$\times10^{15}$\,\cmtw~using vibrationally excited CH$^{13}_3$CN and \chtcn~transitions.
Discrepancies with \citet{2011A&A...525A.151B} probably come from a different distance adopted for the 
source and the fact that we used optically thick lines in our XCLASS model and the rotational diagram.
\citet{2009ApJ...703.1308K}, using only the CH$_3 ^{12}$CN emission, derived a temperature of 323$\pm$105\,K 
and column density of 1$\times10^{15}$\,\cmtw. We find a \chtcn~abundance of 1$\times10^{-8}$. 
With the uncertainties, our results agree with \citet{2009ApJ...703.1308K}.

{\bf I18182} shows OH, Class II \chtoh, and \water~ maser
emission, weak cm continuum emission, and  multiple molecular outflows 
\citep{1998MNRAS.301..640W,2006AJ....131..939Z,2006A&A...454..221Beuther}. 
High-density gas tracers (\chtcn, \chtoh, and HCOOCH$_3$) appear
offset from the\,mm continuum peak, but they are associated with the outflows \citep{2006A&A...454..221Beuther}. 
\citet{2006A&A...454..221Beuther} found gas masses of 47.6 and 12.4\,\mo~from the 1.3\,mm continuum emission using 
dust temperatures of 43 and 150\,K, respectively. They estimated H$_2$ column densities of $5.7\times10^{23}$ and
$1.5\times10^{23}$\,\cmtw~at the same temperatures. Their XCLASS analysis of the \chtcnte~shows 
a rotation temperature of 150\,K and a column density of $3.5\times10^{14}$\,\cmtw, using a single component model.
With the same data and assuming a temperature of 219\,K, we obtained a gas mass of $\sim$21\,\mo~and a H$_2$ 
column density of $3.3\times10^{23}$\,\cmtw. Assuming two components for the XCLASS analysis, we obtain rotation 
temperatures of 219 and 75\,K, and column densities of 7.3$\times10^{15}$ and 8.2$\times10^{13}$\,\cmtw. 
We estimated $X_{\rm CH_3CN} = 2.1\times10^{-8}$.

{\bf G23.01} is a relatively isolated MSFR showing complex OH, \water, and \chtoh~class II maser
emission \citep{1983JRASC..77..257C,1989A&A...213..339F,2011ARep...55..445P} but no free-free emission. Masers are clustered within 2000 AU in a probable disk,
from which an outflow emerges. A 1.3 cm continuum source likely
traces a thermal jet driving the massive CO outflow observed at large
scales \citep{2010A&A...517A..78S}. From the analysis of \chtcn(6-5) transitions \citet{2008ApJ...673..363F} 
found a rotation temperature of $\sim$121\,K and a column density of $4.6\times10^{14}$\,\cmtw, using the RD method.
They found a gas mass of $\sim$380\,\mo~and H$_2$ column density of $3.6\times10^{23}$\,\cmtw~from the 3\,mm continuum 
emission. From 1.3\,mm dust continuum emission, we estimate a gas mass of $\sim$16\,\mo~and H$_2$ column density of 
$1.4\times10^{23}$\,\cmtw. The difference in gas mass with \citep{2008ApJ...673..363F} probably comes from 
our smaller size 
source and higher temperature. With the XCLASS analysis we calculate rotation temperatures of 237 and 58\,K, 
and \chtcn~column densities of $1.5\times10^{17}$ and $1.7\times10^{14}$\,\cmtw, for the compact and extended 
components. We estimated a \chtcn~abundance of 1$\times10^{-7}$.

{\bf G28.20N} is an \hchii~region showing \water, OH, and
\chtoh~maser emission \citep{1995MNRAS.273..328C,2000ApJS..129..159A}. Rotation and probably infall motion of gas
is detected with \nht~ \citep{2005ApJ...631..399S}. From SMA observations of
\chtcn~\citet{2008ApJ...686L..21Q} estimated a rotation temperature of 300\,K, 
column density of $1.6\times10^{16}$\,\cmtw~and \chtcn~fractional abundance of $5\times10^{-9}$, 
by rotation diagrams.
From the 1.3\,mm dust continuum emission, we estimate a gas mass of 33\,\mo~and 
H$_2$ column density of $7.7\times10^{24}$\,\cmtw. Using the XCLASS program, we estimate rotation 
temperatures of 295 and 59\,K, and \chtcn~column densities of $6.2\times10^{16}$ and $2.4\times10^{14}$\,\cmtw, 
for the compact and extended regions, and we calculate $X_{\rm CH_3CN} = 8\times10^{-8}$.

{\bf G31.41} is a prototypical HMC imaged in multiple
high-excitation molecular transitions such as \nht(4,4),
\chtcn(6-5) and (12-11), \chtoh, CH$_3$CCH, and others \citep{1994ApJ...435L.137C,2008ApJ...675..420Araya,1998A&AS..133...29H}.  
\citet{1994ApJ...435L.137C} detected \chtcn(6-5), CH$^{13}_3$CN(6-5), and vibrationally excited \chtcn(6-5),
and estimated a rotation temperature of 200\,K. \citet{1998A&AS..133...29H} observed \chtcn(13-12) and (19-18) 
and report 
temperatures of 149 and 142\,K, respectively, and a column density $>0.5\times10^{14}$\,\cmth. 
\citet{1996A&A...307..599O}, using observations with the IRAM 30\,m of several \chtcn~transitions, estimated a 
rotation temperature of $\sim$140\,K and a column density of $2.3\times10^{17}$\,\cmtw. 
\citet{1998A&A...331..709C} found kinetic temperatures of 250-400\,K toward the cores, 
using the NH$_3$(4,4) line. Recently, with two SMA configurations and IRAM 30m 
observations of \chtcn(12-11) and CH$^{13}_3$CN, \citet{2011A&A...533A..73C} confirmed the existence of a 
velocity gradient, 
explained it as a rotating toroid. Using only the compact SMA configuration data as \citet{2011A&A...533A..73C}, 
we obtain 
a gas mass $\sim$250\,\mo~and H$_2$ column density of $3.5\times10^{24}$\,\cmtw. With the XCLASS analysis 
we calculate 
rotational temperatures of 327 and 95\,K, and column densities of $1.5\times10^{17}$ and $9.1\times10^{13}$\,\cmtw~
for the compact 
and extended regions. We estimate a \chtcn~abundance of $4.2\times10^{-8}$.

{\bf I18566} shows \water, \chtoh~and H$_2$CO maser emission, CS, an outflow 
traced by \nht~and SiO, and weak emission at 3.6\,cm and 2\,cm, probably coming from an 
ionized jet \citep{2007A&A...470..269Z,2005ApJ...618..339Araya,2002A&A...390..289Beuther}. 
This source harbors a 6 cm H$_2$CO maser that flared in 2002 
\citep{2007ApJ...654L..95Araya}. From 43 and 87\,GHz continuum emission, \citet{2007A&A...470..269Z}
estimate a gas mass of $\sim$70\,\mo~for the core. Also, they detected 
significant heating of the \nht~gas (70 K) as a consequence of the outflow. 
From the 1.3\,mm dust continuum we estimate a gas mass of $\sim$16\,\mo~and H$_2$ column density of 
$1.1\times10^{23}$\,\cmtw. Our XCLASS analysis gives rotation temperatures 
of 382 and 110\,K, and \chtcn~column densities of $7.1\times10^{15}$ and $5.1\times10^{13}$\,\cmtw. 
We detected broad linewidths for most of the \chtcn-$K$ components, with an average FWHM
of 8.2\,\kms. We find a \chtcn~abundance of 6.4$\times10^{-8}$.

{\bf G45.07} is a pair of spherical \uchii~regions showing
OH, \water, and \chtoh~maser emission. At least three continuum sources are observed 
in the mid-infrared \citep{2005ApJ...635..452D}. 
\citet{1997ApJ...478..283H} observed CS and CO probably tracing an outflow; 
the \water~ masers are roughly in the same direction as the axis. From the 1.3\,mm dust 
continuum emission, we estimate a gas mass of 172\,\mo~and H$_2$ column density of 
$2.7\times10^{24}$\,\cmtw. The physical parameters obtained from XCLASS are 290 and 82\,K and 
\chtcn~column densities of $5.4\times10^{15}$ and $4.1\times10^{14}$\,\cmtw, 
for the compact and extended regions, respectively.
We estimate a \chtcn~abundance of $1.9\times10^{-9}$.
 
{\bf G45.47} is a MSFR associated with an \uchii~region, multiple molecular lines 
and OH, \water, and \chtoh~maser emission 
\citep{1992A&A...256..618C,2004ApJ...617..384R,1993A&A...276..489O}. 
\citet{1993A&A...276..489O} detected \chtcn(6-5), (8-7), and (12-11) transitions using the IRAM 
30 m, and estimated upper limits of 51\,K for the rotation temperature and 
$2.5\times10^{13}$\,\cmtw~for the column density. 
From NH$_3$(2,2) and (4,4), \citet{1999ApJ...514..899H} estimated a rotation temperature of 
59\,K and column density of $1.8\times10^{17}$\,\cmtw. From the ammonia absorption, they 
suggested that molecular gas is infalling onto the \uchii~region. 
From their molecular line surveys \citet{1998A&AS..133...29H} and \citet{2004ApJ...617..384R} 
report that G45.47 is line-poor; they do not see evidence for a HMC in this field. 
However, the \uchii~region and relatively high luminosity ($\sim$10$^6$\,\lo) suggest a 
more-evolved MSFR. From the 1.3\,mm we estimate a gas mass of $\sim$44\,\mo~and H$_2$ column density of 
$6.7\times10^{23}$\,\cmtw. From the XCLASS analysis we estimate rotation temperatures between 
155\,K and 65\,K, column densities of $6.8\times10^{14}$ and $3.8\times10^{13}$\,\cmtw, and 
$X_{\rm CH_3CN} = 1\times10^{-9}$. Consistent with the molecular line surveys mentioned above, we find
little molecular line emission from this source, compared to the rest of our sample.

{\bf W51e2} is an \uchii~region associated with warm gas and \water,
OH, \nht, and $\rm CH_3OH$ maser emission \citep{1987ApJS...65..193G,1993ApJ...417..645G,
1995ApJ...450L..63Z,1997ApJ...488..241Z,1998ApJ...494..636Z}.
From the 1.3\,mm continuum analysis, \citet{2009ApJ...703.1308K} estimated a gas mass of 
140\,\mo~assuming a dust temperature of 400\,K. They calculated from the \chtcn~a rotation temperature of 
460\,K and column density of $2.1\times10^{16}$\,\cmtw. Using the same data, we estimate 
a gas mass of 95\,\mo~and column density of $3.9\times10^{16}$\,\cmtw. Differences probably come from 
our higher temperature for the dust emission. 
Using the RD method, \citet{1998ApJ...494..636Z} found a \chtcn~column density of 
$2.8\times10^{14}$\,\cmtw~and rotation temperature of 140\,K; they estimated 
a \chtcn~fractional abundance of $5\times10^{-10}$.
From multiple \chtcn~transitions at 3\,mm and 1\,mm, and using a LTE model for each transition, 
\citet{2004ApJ...606..917R} estimated a rotation temperature of 153\,K and column density 
of  $3.8\times10^{16}$\,\cmtw. They calculated an H$_2$ column density of $8.3\times10^{22}$\,\cmtw~and 
\chtcn~fractional abundance, $X_{\rm CH_3CN} = 4.6\times10^{-7}$. With XCLASS we estimate rotation 
temperatures of 458 and 118\,K, and \chtcn~column densities of $2.8\times10^{17}$ and $1.0\times10^{15}$ 
\cmtw, for the compact and extended components. We calculated $X_{\rm CH_3CN} = 7\times10^{-8}$.
The discrepancies with \citet{2004ApJ...606..917R} probably arise from 
differences in the methods used to estimate temperatures and H$_2$ and \chtcn~column densities.
The much lower abundances reported by \citet{1998ApJ...494..636Z} are a direct consequence of the 
much lower column density that they report from \chtcn.

{\bf W51e8} is a MSFR located to the south of W51e2, and is associated with H$_2$O and OH maser 
emission and multiple molecular lines such as HCO$^+$, NH$_3$, and 
\chtcn~\citep{1997ApJ...488..241Z,1998ApJ...494..636Z}. 
Observing with the Nobeyama Millimeter Array at 2\,mm, \citet{1998ApJ...494..636Z} 
detected molecules such as CS, CH$_3$OCH$_3$, HCOOCH$_3$, and \chtcn. From the latter, they 
estimated a rotation temperature of 130\,K and \chtcn~column density of $2.0\times10^{14}$\,\cmtw. 
\citet{2009ApJ...703.1308K} estimated a dust-derived mass of 82\,\mo, assuming an average temperature of 
400\,K. They estimated a rotation temperature of 350\,K and a \chtcn~column density of $8\times10^{15}$\,\cmtw~
through rotation diagrams. With the same method, \citet{2004ApJ...606..917R} (W51e1 in their nomenclature) 
estimated a rotation temperature of 123\,K, column density of $1.4\times10^{16}$, and fractional 
abundances of 1.3\x10$^{-7}$. With the set of data at 1.3\,mm of \citet{2009ApJ...703.1308K}, 
we calculate an H$_2$ gas mass of 86\,\mo~and a column density of $5.8\times10^{24}$\,\cmtw.
Using the XCLASS program we estimate rotation
temperatures of 384 and 85\,K, and column densities of $1.8\times10^{17}$ and $1.8\times10^{14}$\,\cmtw, 
for the compact and extended regions, respectively. For the compact component, we estimate $X_{\rm CH_3CN}$ of 3.4\x10$^{-8}$. 
As in the case of W51e2, the differences with \citet{2004ApJ...606..917R} come from the methods used to obtain 
the physical parameters. 

\section{DISCUSSION}\label{discussion}

\subsection{Mass and density from 1.3\,mm continuum}\label{masscont} 

The gas masses estimated from the 1.3\,mm emission range from 7 to 375\,\mo, and for column densities 
from 1.0\x10$^{23}$ to 6.7\x10$^{24}$\,\cmtw. The median value for the mass is 21\,\mo~and for 
the column densities 8\x10$^{23}$\,\cmtw. Using the physical size of a deconvolved 
beam, we obtained H$_2$ number densities from 8\x10$^{5}$ to 1.4\x10$^{8}$\,\cmth, assuming that the gas is 
distributed uniformly.

Except for IRAS17233, all sources are more massive than $\sim$10\,\mo, which is 
commonly adopted as the lower limit for the gas mass of a HMC \citep[][]{2005IAUS..227...59C}.

The size of the dusty structures 
goes from $\sim$3500 to 20500 AU (0.017 to 0.10\,pc). In general, the 
dust-emission structures shown in Figures \ref{lines1} and \ref{lines2} have similar 
sizes, gas masses and column densities to other HMCs \citep[e.g.,][]{2011A&A...525A.151B}.

From Table \ref{mmcont} we see that the estimated thermal dust contribution 
to the total 1.3\,mm flux ranges from 8\% in G5.89, to $>$98\% in sources W3TW, I16547, 
I17233, G8.68, I18182, G23.01, G31.41, and I18566. The latter sources show a high fraction of dust 
emission because they have essentially no free-free emission from ionized gas. 
Since even this sub-group has luminosities $>10^4$\,\lo~(corresponding to an early B-type star) we would expect a 
more substantial amount of ionization. Possible reasons for the lack of such ionization include young sources in 
early evolutionary stages, very high mass accretion rates, or the presence of a stellar cluster whose 
luminosity is dominated by late-B-type stars.
Sources such as G10.47, G31.41, G45.07, G45.47, and W51e2 clearly show \uchii~regions embedded 
in the dusty molecular gas, indicating much higher levels of ionization.

One of the implicit assumptions in Equations \ref{eq:mgasfin} and \ref{eq:ngasfin}
is that the gas and dust are well-coupled. As a test, we calculate the gas-dust relaxation time, $t_{\rm gd}$. 
We follow \citet{2006ApJ...639..975C}, who estimated the time-scale necessary for thermal 
coupling of the dust and gas, and obtained $t_{\rm gd}=2.5\times10^{16} / n_{\rm H}$, in seconds, 
where $n_{\rm H}$ is the number density of H nuclei in\,\cmth. For the values of $n_{\rm H_2}$ in Table \ref{mmcont}, 
the gas-dust relaxation time ranges from 3 to 500\,yr. Thus, $t_{\rm gd}$ is much shorter 
than the expected HMC lifetime.

With respect to the LTE condition, the \chtcn~critical density is $\sim10^6$\,\cmth~for the $J_{\rm up}=12$ 
transition \citep{2010ApJ...713.1192W}. 
We therefore conclude that the LTE approximation is valid and the rotation temperature of \chtcn~can be taken 
as the kinematic temperature of the H$_2$ gas. 

\subsection{Spatial distribution with respect to the \chtcn}\label{spatial}

In Figures \ref{lines1} to \ref{lines3} we compare the spatial distribution 
of the 1.3\,mm continuum emission with the velocity-integrated emission (moment 0) of 
$K$--lines 3, 5 and 7 of \chtcn. These $K$--components trace kinetic temperatures of 133\,K, 247\,K, and 
418\,K, respectively. For G5.89 and I18182 we use the $K$--lines 3, 5 and 6, owing to the lack of $K=7$
emission.

At the resolution and sensitivity of our data, the 1.3\,mm continuum emission and the hot \chtcn~emission
coincide closely and have similar spatial extents for most of the sources. 
Notably, G5.89 has a more complicated morphology.

The close spatial coincidence between dust and molecular gas can be explained if embedded protostars are
heating the dust grains, thus evaporating the ice mantles, and subsequent gas-phase chemical reactions produce species
such as \chtcn. This scenario has been observed toward HMCs, specially with nitrogen-bearing molecules 
\citep[e.g.,][]{2010ApJ...711..399Q}. In Section 5.4 we will explore various scenarios to explain the molecular abundances. 

We note that the displacement of some \chtcn~lines from the continuum peak for sources such G5.89, G10.47, G28.20N and G45.47 
(see Figures \ref{lines1} to \ref{lines3}) could be due to factors such as multiple star forming points, 
unresolved continuum sources or molecular gas heated externally. 
Since, higher-$K$ transitions should be excited in denser, hotter and probably more compact regions,
clumpy cores, with different physical conditions, are an alternative explanation for the displacements.
Sub-arcsecond observations will be necessary to test these alternative explanations.

 \subsection{Temperature, density, and virial mass}\label{massmol}  

The \chtcn~analysis, using the rotational diagrams and XCLASS, indicates
high temperatures and densities for all sources. However,
from Figure \ref{RDs}, large deviations are clearly seen to the linear fit for some of the $K=3,6$-lines, 
toward sources such IRAS17233, G10.47, G10.62, G31.41, IRAS16547, and IRAS18566; this indicates that 
these lines are optically thick and hence there is probably a mixture of optically thin and thick lines in our 
spectra. This occurs because the $K=0,3,6,...$, ladders are doubly degenerate compared to the $K=1,2,4,5,...$ ladders. 
Thus the former lines will have higher optical depths than the latter. As we mention below, the homogeneous 
assumption of the RD method probably is inadequate.   

Moreover, Figures \ref{xclass1} and \ref{xclass2} show similar brightness temperatures of 
low $K$-ladders ($K=0-4$), including the $K=3$-line, toward some sources (e.g., IRAS17233, G28.20N, G10.47, 
G31.41, W51e2, and W51e8). This confirms that these lines are optically thick. 

We find good agreement between the observed spectra and the synthetic models using XCLASS, with a two-component
model. This result, along with the close coincidence between the molecular gas and dust 
emission, suggest that most of the HMCs are internally heated.

We consider the two-component model to be more realistic than a single homogeneous structure. The 
reality is probably even more complex; \citet[][]{2010A&A...509A..50C,2011A&A...533A..73C}, for example, 
report sub-arcsecond observations that indicate gradients in both temperature and density.
We note that the two-component model overestimate the $K=3$-line for some sources, suggesting that even 
this model is too simplistic.

We summarize our results from the XCLASS program as follows. For the extended component, the temperature ranges from  
40 to 132\,K, with an average of 85\,K, and median of 82\,K; the mean size is 0.10\,pc; the average column density 
$N_{\rm CH_3CN}=2.4$\x10$^{14}$\,\cmtw, with a median of $1.7$\x10$^{14}$\,\cmtw. 
For the compact component, the temperature ranges from 122 to 485\,K, with an average of 303\,K, and median of 295\,K; 
the mean size is 0.02\,pc; the average column density $N_{\rm CH_3CN}=8.5$\x10$^{16}$\,\cmtw, with a median of 
$2.2$\x10$^{16}$\,\cmtw. 

In Tables 3 and 5 we present \mgas~ and $M_{\rm vir}$, which provide information about the stability and 
structure of the HMCs. The ratio of the virial mass to gas mass, $M_{\rm vir}$/\mgas~, is greater than unity for 
all sources but G10.47 (see Figure 10). The ratio ranges from 29.6 to 1.1, with an average 
of 8.3. This suggests that the cores are not in virial equilibrium and a traditional interpretation 
of this result is that the HMCs are expanding, since E$_{\rm k}$ $>$2E$_{\rm g}$. However, another interpretation 
is that these cores are still collapsing, as suggested by recent models of molecular cloud formation and 
evolution \citep[see][]{2011MNRAS.411...65B}. This possibility will be explored in a future work.

\subsection{Fractional Abundances}

Adopting the H$_2$ column densities from the 1.3\,mm continuum emission,
we find $\rm CH_3CN$ abundances, $X_{\rm CH_3CN}$, from $\sim$1\x10$^{-9}$ to $\sim$2\x10$^{-7}$ toward the 
hot-inner components. These results span a range of fractional abundances 
that agrees with other estimates toward HMCs, such as the Orion hot core with 10$^{-10}$-10$^{-9}$ 
\citep[][]{1994ApJ...422..642Wilner}, G20.08N with 5\x10$^{-9}$ to 2\x10$^{-8}$ \citep[][]{2009ApJ...706.1036G}, 
Sgr B2(N) with $\sim$3\x10$^{-8}$ \citep[][]{2000ApJS..128..213N}, and W51e8 and W51e2 with 1.3\x10$^{-7}$ 
and 4.6\x10$^{-7}$, respectively \citep{2004ApJ...606..917R}.
 
At the high dust temperatures of the compact components (T$>$122 K), most of the organic molecules 
are probably evaporated from the grain mantles and incorporated into the gas phase \citep{2009ARA&A..47..427H}.  
Moreover, these temperatures are high enough to form many new organic species by chemical reactions 
--- if the chemical time scales are short enough. For any given molecular species, one can ask if it was
formed 1) in a dense, cold gas phase prior to any protostelar object, 2) on the grains by surface reactions 
or 3) in gas phase processes after evaporation \citep[see the review by][]{2009ARA&A..47..427H}. In the case 
of \chtcn~all three of these scenarios have been studied 
using both chemical models and observations \citep[e.g.,][]{1995LNP...459..291M,1998FaDi..109..205O,2001ApJ...546..324R,2010ApJ...713.1192W}. 

Observations toward cold dense gas show \chtcn~abundances of $\sim10^{-10}$ (Ohishi \& Kaifu 1998)
 i.e., substantially lower than the abundances that we find. 
Thus, the scenario in which \chtcn~is formed in a cold gas phase, then adsorbed by dust grains and 
released by heating from young protostars appears not to contribute to the high abundances observed 
toward these HMCs.

Alternatively, formation of \chtcn~on grain surfaces and/or in the hot gas after evaporation 
of \ca parent\cc~species represent better possibilities. The former process tends to underestimate 
the final abundances \citep[e.g.,][]{Caselli+1993}, while the chain of mantle-surface-gas 
reactions yields good agreement with HMC abundances at times $>10^5$\,yr \citep[e.g.,][]{1993MNRAS.263..589H}.
Moreover, if we consider the grain-surface process as the main path to form \chtcn, the abundances 
toward the inner regions of all HMCs should be very similar, because the high temperatures would sublimate 
most of the ice on the dust \citep[e.g.,][]{2004MNRAS.354.1141V,2009ARA&A..47..427H}

Chemical models, suggest that \chtcn~is synthesized from \nht~and HCN, once the ammonia 
is released from ice mantles, and via the ion-molecule reaction of CH$^+_3$ + HCN and the radiative association reaction of 
CH$_3$ with CN \citep[][]{1992ApJ...399L..71C}. The process occurs once the dusty regions reach temperatures $>$100\,K 
\citep[e.g.,][]{2001ApJ...546..324R,2009ARA&A..47..427H}. Moreover, at temperatures $>$300\,K the environment is optimal to form 
\chtcn~from the parent nitrogenated species HCN and \nht \citep[][]{2001ApJ...546..324R}. 
For example, \citet[][]{2002A&A...389..446D} found that the HCN abundance increases with temperature, and 
at T $>$ 200 the formation of HCN proceeds quickly. This temperature dependence in the reactions 
of N-bearing molecules, including HCN and \chtcn, has been observed in other chemical models and HMCs 
\citep[][]{2001ApJ...546..324R}.

In Figure \ref{plots2} we plot $X_{\rm CH_3CN}$ versus rotation temperature estimated for the compact components with XCLASS, 
and our RD results. Uncertainties of the hot components from XCLASS program were used to estimate uncertainties 
in gas masses and column densities from the continuum emission. We observe that fractional abundances increase with higher temperatures. 
This result can be understood with the chemical scenario in which \chtcn~molecules mainly form in a hot gas phase 
and its production is optimized at higher temperatures. A similar 
dependence between \chtcn~abundance and temperature was detected toward Orion-KL and the Compact Ridge 
\citep[][]{1994ApJ...422..642Wilner,2010ApJ...713.1192W}. High angular resolution, multi-species molecular line 
observations with ALMA would provide valuable constrains for chemical models to confirm this hypothesis.

\subsection{HMC and \uchii~regions}

From observations and theoretical models, HMCs have been proposed as the cradle of massive stars. 
Once massive stars produce enough UV photons, the surrounding atomic and molecular material will be 
ionized, forming \uchii~regions. In our sample there are five HMCs (41\%) with little or no centimeter 
free-free emission. They could represent young objects on the verge of becoming \uchii~regions.

The evolutionary sequence of HMCs and the time scales involved are not fully understood. 
Using the ratio of the number of HMCs to the number of \uchii~regions, 
\citet[][]{2001ApJ...550L..81Wilner} and \citet[][]{2005ApJ...624..827F}
estimated that HMCs live at least 25\% of the \uchii~region lifetime, i.e., some $10^4$\,yr. 
Similarly, \citet[][]{2000prpl.conf..299Kurtz} estimated a lifetime between 1.9\x10$^3$ and 5.7\x10$^4$\,yr,
based in the number of HMC and \uchii~regions known at that time.

On the other hand, chemical models require $10^4-10^5$\,yr to reach the chemical richness observed in HMCs 
(\eg~Charnley et al. 1992; Herbst \& van Dishoeck 2009). 
If the chemical models of \citet[][]{2001ApJ...546..324R} are correct, chemically rich HMCs 
should have lifetimes between $10^4$ and $10^5$\,yr, and hence a larger ratio of HMCs to \uchii~
would be expected. Although we are approaching a complete census of \uchii~regions \citep[][]{2013ApJS..205....1P}, 
the statistics of HMCs are not yet known with sufficient accuracy to constrain the 
chemical models.

\section{Conclusions}\label{conclus}

We studied 17 hot molecular cores in the \chtcnte~lines and the 1.3\,mm continuum. 
The sources were observed with the SMA at $\sim$220\,GHz, with either the compact or extended configuration. 

From the 1.3\,mm continuum, we detected dusty structures with physical sizes of 0.01--0.1\,pc, 
gas masses of 7--375\,\mo, and column densities of 0.1--6.7\x10$^{24}$\,\cmtw. 
The continuum emission coming from dust ranges from 8\% to 98\% 
of the total flux. 

All 17 sources show multiple molecular lines but different molecular richness. 
All sources show five or more $K$-components of \chtcnte.
Some spectra showed emission up to the $K=8$-component, which traces gas at $\sim$525\,K.

Based on these emission lines we estimated rotational temperatures, column densities, and fractional 
abundances, using both rotation diagrams and the XCLASS program that generates synthetic spectra. 
From the rotation diagram method we find temperatures from 90 to 500\,K, and column densities from 2.5\x10$^{13}$ 
to 2.5\x10$^{16}$\,\cmtw. With XCLASS we find temperatures from 40 to 132\,K and column densities from 1.6\x10$^{13}$ 
to 1.0\x10$^{15}$\,\cmtw~for the extended component, and temperatures from 122 to 485\,K and column densities from 6.8\x10$^{14}$ 
to 5.1\x10$^{17}$\,\cmtw~for the compact component.

We used the rotation temperatures estimated with XCLASS to derive the gas mass from the 1.3\,mm continuum. 
With the multiple $K-$lines of \chtcn~we find a good fit between observed and synthetic spectra for the 
two-component XCLASS model; this result, and a close spatial coincidence between the molecular gas and the 
continuum emission, suggest that most of these HMC are internally heated. Sub-arcsecond observations 
are necessary to explore their structure in greater detail.

The fractional abundance of \chtcn~toward the hot-inner components shows a marked increase with temperature. 
This can be understood if we consider that \chtcn~molecules form in the hot gas phase when 
parent N-bearing species, such as \nht, are evaporated from grain mantles. 

\acknowledgments

This work has received partial support from grants UNAM/DGAPA project IN101310 to SK, and CONACYT fellowship 
to VH-H. VH-H thanks Roberto Galvan-Madrid for comments and helpful discussion. 
We thank the anonymous referee for comments and suggestions that significantly improved the manuscript. 
               
\bibliography{../myrefs/myref}

\newpage

\begin{deluxetable*}{lcccccccc}
\tabletypesize{\scriptsize}
\tablecaption{Observed Sources \label{sources}}
\tablewidth{0pc}
\tablehead{
\colhead{Source} & \colhead{Short} & \colhead{R.A.}    & \colhead{Dec.}    & \colhead{V$_{\rm lsr}$} & \colhead{Distance} & \colhead{$L$}         &                  &  \colhead{}   \\      
\colhead{Name}   & \colhead{Name}  & \colhead{(J2000)} & \colhead{(J2000)} & \colhead{(km/s)}    & \colhead{(kpc)}    & \colhead{($10^5$\lo)} & \colhead{\uchii} &  \colhead{Refs} }    
 \startdata                                                                                                               
W3(OH)           & W3OH      & 02 27 03.9   & +61 52 24   & --48.0    & 2.0      & 0.10        & Y & 1     \\         
W3(H2O)TW        & W3TW      & 02 27 04.8   & +61 52 24   & --48.0    & 2.0      & 0.30        & N & 1     \\         
IRAS 16547--4247 & I16547    & 16 58 17.3   & --42 52 07  & --30.0    & 2.9      & 0.62        & N & 2,3     \\         
IRAS 17233--3606 & I17233    & 17 26 42.8   & --36 09 17  & --03.4    & 1.0      & 0.14        & Y & 4,5     \\         
G5.89--0.37      & G5.89     & 18 00 30.4   & --24 04 00  & +10.0     & 3.0      & 1.50        & Y & 1     \\         
G8.68--0.37      & G8.68     & 18 06 23.4   & --21 37 10  & +37.2     & 4.8      & 0.20        & N & 6,7   \\         
G10.47+0.03      & G10.47    & 18 08 38.2   & --19 51 50  & +68.0     & 8.5      & 7.00        & Y & 1   \\         
G10.62--0.38     & G10.62    & 18 10 28.7   & --19 55 49  & --03.0    & 5.0      & 9.20        & Y & 1   \\         
IRAS 18182--1433 & I18182    & 18 21 09.0   & --14 31 49  & +59.1     & 3.6      & 0.19        & N & 1   \\         
G23.01--0.41     & G23.01    & 18 34 40.3   & --09 00 38  & +77.0     & 4.6      & 1.00        & N & 1   \\         
G28.20--0.04N    & G28.20N   & 18 42 58.1   & --04 13 57  & +95.6     & 5.7      & 1.60        & Y & 8,9   \\         
G31.41+0.31      & G31.41    & 18 47 34.3   & --01 12 46  & +96.5     & 7.9      & 2.50        & Y & 10,11   \\         
IRAS 18566+0408  & I18566    & 18 59 09.8   & +04 12 13   & +85.0     & 6.7      & 0.60        & Y & 12   \\         
G45.07+0.13      & G45.07    & 19 13 22.0   & +10 50 53   & +60.0     & 8.0      & 11.0        & Y & 1   \\         
G45.47+0.05      & G45.47    & 19 14 25.6   & +11 09 25   & +59.0     & 8.0      & 11.0        & Y & 1   \\         
W51e2            & W51e2     & 19 23 44.0   & +14 30 35   & +55.0     & 5.4      & 15.0        & Y & 1   \\         
W51e8            & W51e8     & 19 23 43.9   & +14 30 28   & +55.0     & 5.4      & 15.0        & Y & 1           
\enddata
\tablecomments{Units of right ascension are hours, minutes, and seconds, and for declination are degrees, arcminutes, 
and arcseconds. Positions, V$_{\rm lsr}$, distances, luminosities, and \uchii~regions are taken from the cited references.} 
\tablerefs{ 
(1) \citep[][]{2014arXiv1401.5377R}; (2) \citet{2008AJ....135.2370R}; (3) \citet{2009ApJ...701..974F}; 
(4) \citet{2004A&A...426...97F}; (5) \citet{2011A&A...530A..12L}; (6) \citet{2006MNRAS.367..553P}; 
(7) \citet{2011ApJ...726...97L}; (8) \citet{2005ApJ...631..399S}; (9) \citet{2008ApJ...686L..21Q};
(10) \citet{2008A&A...486..191P}; (11) \citet{2010A&A...509A..50C}; (12) \citet{2007A&A...470..269Z} }
\end{deluxetable*}


\begin{deluxetable*}{lcccccccccc}
\tabletypesize{\scriptsize}
\tablewidth{0pc}
\tablecaption{Observational Parameters \label{obspar}}
\tablehead{
\colhead{Source}&\colhead{Observation} & \colhead{Frequency}      & \colhead{Spectral}  & \multicolumn{3}{c}{Calibrators} & \colhead{$S_{\nu}$ of gain}& \multicolumn{2}{c}{Synthesized beam} & \colhead{Published}  \\ \cline{5-7} \cline{9-10}
\colhead{Name} &\colhead{Epoch}&\colhead{range of LSB}&\colhead{resolution}&\colhead{Bandpass}&\colhead{Gain}&\colhead{Flux}&\colhead{calibrators\tablenotemark{a}}&\colhead{FWHM}&\colhead{PA}&\colhead{data} \\
\colhead{}     &\colhead{}  & \colhead{(GHz)} & \colhead{(MHz)} & \colhead{} & \colhead{} & \colhead{} & \colhead{(Jy)} & \colhead{(arcsec)} & \colhead{(deg)} & \colhead{} }  
\startdata                                                                                                                                        
W3OH      & 2004 Oct 24 & 219.50--221.48 & 0.812   & 0359+509 & 0102+584      & Uranus      & 1.41  & 2.97\x1.93 & +69.1 & \nodata     \\ 
          &             &                &         &          & 0359+509      &             & 3.10  &            &       & \colhead{}  \\                    
W3TW      & 2004 Oct 24 & 219.50--221.48 & 0.812   & 0102+584 & 0102+584      & Uranus      & 1.41  & 2.97\x1.93 & +69.1 & \nodata     \\ 
          &             &                &         &          & 0359+509      &             & 3.10                       & \colhead{}  \\                                 
I16547    & 2006 Jun 06 & 219.21--221.19 & 0.812   & 3C273    & 1745-290      & Uranus      & 3.09  & 1.97\x1.18 & --6.3 & \nodata     \\ 
          &             &                &         &          & 1604-446      &             & 1.32                       & \colhead{}  \\           
I17233    & 2007 Apr 10 & 219.45--221.43 & 0.406   & 3C454.3  & 1626-298      & Callisto    & 1.22  & 4.85\x2.14 & +32.1 & 1           \\ 
          &             &                &         &          & 1713-269      &             & 0.30                       &             \\     
G5.89     & 2008 Apr 18 & 219.37--221.34 & 0.406   & 3C454.3  & 1921--293     & Uranus      & 1.11  & 3.25\x2.00 & +60.4 & 2           \\ 
          &             &                &         & 3C273    & 1733-130      &             & 0.71                       &             \\                             
G8.68     & 2008 Sep 17 & 220.28--222.27 & 0.406   & 3C454.3  & 1911--201     & Uranus      & 1.89  & 3.71\x2.79 & +10.8 & 3           \\ 
          &             &                &         &          & 1733-130      &             & 2.70                       &             \\ 
G10.47    & 2008 Jun 21 & 220.24--222.22 & 0.406   & 3C279    & 1733--130     & Uranus      & 1.39  & 3.26\x1.91 & +63.9 & \nodata     \\ 
          &             &                &         & 3C454.3  & 1911-201      &             & 1.05                       & \colhead{}  \\                      
G10.62    & 2009 Jan 31 & 220.32--222.30 & 0.406   & 3C454.3  & 1733-130      & Uranus      & 2.07  & 5.34\x2.95 & --0.9 & \nodata     \\ 
          &             &                &         &          & \nodata       &             &\nodata&            &       &             \\    
I18182    & 2004 Apr 30 & 219.42--221.07 & 0.812   & 3C273    & 1733-130      & Uranus      & 1.48  & 3.84\x2.59 & +15.9 & 4           \\
          &             &                &         &          & 1908-201      &             & 1.64                       &             \\                 
G23.01    & 2010 Apr 28 & 216.90--220.88\tablenotemark{b}& 0.812 & 3C273& 1743-038 & Uranus & 0.97  & 3.52\x3.16 & --43.7& \nodata     \\ 
          &             &                &         & 3C454.3  & 1911-201      &             & 1.55                       & \colhead{}  \\                      
G28.20N   & 2008 Jun 21 & 220.25--222.22 & 0.406   & 3C279    & 1733--130     & Uranus      & 1.39  & 3.31\x1.60 & --72.0& \nodata     \\ 
          &             &                &         & 3C454.3  & 1911-201      &             & 1.05                       & \colhead{}  \\     
G31.41    & 2007 Jul 09 & 219.30--221.30 & 0.406   & 3C273    & 1751+096      & Uranus      & 1.59  & 3.53\x1.70 & +66.0 & \nodata     \\ 
          &             &                &         &          & 1830+063      &             & 0.46                       & \colhead{}  \\       
I18566    & 2007 Jul 09 & 219.30--221.30 & 0.812   & 3C273    & 1751+096      & Uranus      & 1.59  & 3.48\x1.65 & +66.7 & \nodata     \\           
          &             &                &         &          & 1830+063      &             & 0.46                       & \colhead{}  \\              
G45.07    & 2007 Apr 13 & 219.45--221.43 & 0.812   & 3C279    & 1751+096      & Callisto    & 8.88  & 3.31\x1.60 & +78.7 & \nodata     \\ 
          &             &                &         &          & 1925+211      &             & 1.65                       & \colhead{}  \\    
G45.47    & 2008 Jun 30 & 219.15--221.13 & 0.406   & 3C454.3  & 1925+211      & Titan       & 0.63  & 3.40\x1.68 & +77.2 &  \nodata    \\
          &             &                &         &          & 1911-201      &             & 1.41                       & \colhead{}  \\             
W51e2     & 2005 Sep 01 & 220.25--222.23 & 0.406   & 3C454.3  & 1751+096      & Uranus      & 1.41  & 1.47\x0.83 & -86.9 & 5           \\ 
          &             &                &         &          & 2025+337      &             & 0.74                       &  \colhead{} \\                       
W51e8     & 2005 Sep 01 & 220.25--222.23 & 0.406   & 3C454.3  & 1751+096      & Uranus      & 1.41  & 1.47\x0.83 & -86.0 & 5           \\ 
          &             &                &         &          & 2025+337      &             & 0.74                       & \colhead{}                                  
\enddata
\tablecomments{(a) We estimated bootstrapped flux for gain calibrators with an uncertainty of 15\%--20\%. (b) A single sideband of 4\,GHz bandwidth.}
\tablerefs{1: \citet{2011A&A...530A..12L}; 2: Su et al. (2009); 3: Longmore et al. (2011); 4: Beuther et al. (2006); 5: Klaassen et al. (2009)}
\end{deluxetable*}


\begin{deluxetable*}{lcccccccccc}
\tabletypesize{\scriptsize}
\tablewidth{0pc}
\tablecaption{1.3\,mm continuum results \label{mmcont}}
\tablehead{ 
\colhead{}& \colhead{R.A.\tablenotemark{a}} & \colhead{Dec.\tablenotemark{a}} &\colhead{$S_{\rm \nu}^{\rm Peak}$}& \colhead{$S_{\nu}^{\rm Total}$}&\colhead{$S_{\rm \nu}^{\rm dust}$}&\colhead{$\theta_{\rm s}$\tablenotemark{b}} & \colhead{Physical Size\tablenotemark{c}}&\colhead{\mgas~\tablenotemark{d}}&\colhead{\Nhtw~} & \colhead{$n_{\rm H_2}$} \\ 
\colhead{Source}&\colhead{(J2000)}&\colhead{(J2000)}&\colhead{(Jy/beam)}&\colhead{(Jy)}&\colhead{(Jy)}&\colhead{(arcsec)}&\colhead{(10$^{-3}$\,pc)}&\colhead{(\mo)}&\colhead{(10$^{24}$ cm$^{-2}$)}&\colhead{(10$^{7}$ cm$^{-3}$)} } 
\startdata                                                                                                                                                        
 W3OH      & 02 27 03.862 &  +61 52 24.60 & 2.42$\pm$0.089 & 3.85$\pm$0.77  &  1.37$\pm$0.27 & 3.2\x2.0 &  30.9 \x 19.6 &  19 &  1.74     &   5.00  \\       
 W3TW      & 02 27 04.611 &  +61 52 24.74 & 1.23$\pm$0.055 & 2.67$\pm$0.53  &  2.66$\pm$0.53 & 3.1\x2.0 &  30.5 \x 18.8 &  12 &  1.18     &   3.49  \\       
 I16547    & 16 58 17.242 & --42 52 07.97 & 0.20$\pm$0.029 & 1.58$\pm$0.32  &  1.57$\pm$0.31 & 4.0\x3.0 &  56.8 \x 41.6 &  21 &  0.48     &   0.70  \\       
 I17233    & 17 26 42.480 & --36 09 17.66 & 2.23$\pm$0.132 & 5.90$\pm$1.18  &  5.88$\pm$1.18 & 4.7\x2.6 &  22.8 \x 12.9 &   7 &  1.35     &   5.57  \\       
 G5.89     & 18 00 30.428 & --24 04 01.62 & 2.15$\pm$0.136 & 7.72$\pm$1.54  &  0.69$\pm$0.14 & 4.5\x3.5 &  65.0 \x 51.5 &  16 &  0.26     &   0.32  \\       
 G8.68     & 18 06 23.492 & --21 37 10.64 & 0.17$\pm$0.010 & 0.40$\pm$0.08  &  0.39$\pm$0.08 & 4.6\x3.0 & 106.8 \x 68.6 &  14 &  0.10     &   0.08  \\       
 G10.47    & 18 08 38.238 & --19 51 50.21 & 3.85$\pm$0.137 & 5.04$\pm$1.00  &  4.90$\pm$0.98 & 1.6\x1.1 &  65.5 \x 46.6 & 375 &  6.69     &   8.61  \\       
 G10.62    & 18 10 28.687 & --19 55 49.17 & 3.70$\pm$0.178 & 7.07$\pm$1.41  &  4.45$\pm$0.89 & 5.0\x3.4 & 120.7 \x 81.9 & 116 &  0.63     &   0.37  \\       
 I18182    & 18 21 09.128 & --14 31 50.56 & 0.39$\pm$0.025 & 0.84$\pm$0.17  &  0.83$\pm$0.17 & 3.4\x3.2 &  60.4 \x 56.5 &  21 &  0.33     &   0.41  \\       
 G23.01    & 18 34 40.297 & --09 00 38.19 & 0.20$\pm$0.009 & 0.42$\pm$0.08  &  0.41$\pm$0.08 & 4.0\x2.9 &  90.3 \x 63.8 &  16 &  0.14     &   0.13  \\       
 G28.20N   & 18 42 58.112 & --04 13 57.56 & 0.75$\pm$0.032 & 1.25$\pm$0.25  &  0.70$\pm$0.14 & 1.9\x1.6 &  53.0 \x 43.9 &  33 &  0.77     &   1.13  \\       
 G31.41    & 18 47 34.334 & --01 12 45.85 & 2.09$\pm$0.070 & 3.06$\pm$0.61  &  3.05$\pm$0.61 & 2.0\x1.3 &  77.7 \x 49.8 & 251 &  3.53     &   4.01  \\       
 I18566    & 18 59 10.001 &  +04 12 15.46 & 0.13$\pm$0.009 & 0.32$\pm$0.06  &  0.31$\pm$0.06 & 3.0\x2.4 &  99.3 \x 78.3 &  16 &  0.11     &   0.08  \\       
 G45.07    & 19 13 22.073 &  +10 50 53.41 & 1.92$\pm$0.085 & 2.92$\pm$0.58  &  1.96$\pm$0.39 & 1.8\x1.3 &  69.4 \x 49.6 & 173 &  2.73     &   3.29  \\       
 G45.47    & 19 14 25.679 &  +11 09 25.54 & 0.35$\pm$0.024 & 0.63$\pm$0.13  &  0.40$\pm$0.08 & 2.6\x1.5 &  95.9 \x 55.2 &  66 &  0.67     &   0.65  \\       
 W51e2     & 19 23 43.947 &  +14 30 34.88 & 1.82$\pm$0.140 & 5.27$\pm$1.05  &  4.14$\pm$0.83 & 1.5\x1.4 &  37.3 \x 35.3 &  96 &  4.00     &   7.70  \\       
 W51e8     & 19 23 43.883 &  +14 30 27.82 & 1.36$\pm$0.107 & 3.12$\pm$0.62  &  2.96$\pm$0.59 & 1.4\x1.0 &  35.8 \x 24.9 &  86 &  5.27     &  12.44         
\enddata
\tablecomments{(a) Positions of the 1.3\,mm continuum peak emission.  (b) Deconvolved sizes from Gaussian fit. (c) Sizes at distances in Table 1. 
(d) Gas mass derived from the estimated 1.3\,mm continuum dust emission ($S_{\rm \nu}^{\rm dust}$) assuming the temperature of the \chtcn~gas in the compact component.}
\end{deluxetable*}


\begin{deluxetable*}{lccccccccccc}
\tabletypesize{\scriptsize}
\tablewidth{0pc} 
\tablecaption{Observed Line Parameters of \chtcnte \label{linedata}}
\tablehead{
\colhead{}      &\colhead{$V_{\rm LSR}$\tablenotemark{a}}&\colhead{$\Delta V$\tablenotemark{a}}& \multicolumn{9}{c}{$\int T dv$ (K\,\kms)}  \\ \cline{4-12}
\colhead{Source}&\colhead{(\kms)}&\colhead{(\kms)}&\colhead{$K=0$}&\colhead{$K=1$}&\colhead{$K=2$}&\colhead{$K=3$}  & \colhead{$K=4$} & \colhead{$K=5$}&\colhead{$K=6$} &\colhead{$K=7$} &\colhead{$K=8$} }    
 \startdata      
W3OH            & -46.2 &  5.3  &  21.5$\pm$0.5  &  32.2$\pm$0.7  &  22.8$\pm$0.4  &  24.1$\pm$0.4   &   9.4$\pm$0.6   &   4.8$\pm$1.1  & 2.2$\pm$1.2    &  \nodata       & \nodata        \\
W3TW            & -49.7 &  6.5  &  35.3$\pm$18.9 &  58.5$\pm$19.8 &  40.1$\pm$3.0  &  39.1$\pm$0.5   &  25.9$\pm$4.6   &  22.2$\pm$5.4  & 13.8$\pm$13.0  &  4.3$\pm$3.1   & \nodata        \\
I16547          & -31.9 &  7.6  &  81.6$\pm$3.8  &  69.8$\pm$2.5  &  66.2$\pm$2.9  &  79.4$\pm$7.3   &  51.4$\pm$5.4   &  30.7$\pm$3.7  & 12.0$\pm$4.2   &  \nodata       & \nodata        \\
I17233          & -3.51 &  9.8  & 201.3$\pm$15.5 & 241.0$\pm$0.6  & 230.9$\pm$1.8  & 269.9$\pm$3.1   & 195.2$\pm$0.7   & 119.9$\pm$10.9 & 157.2$\pm$9.0  &  89.5$\pm$0.6  & 46.8$\pm$2.0   \\
G5.89           &  9.69 &  3.9  &   2.4$\pm$0.4  &   4.3$\pm$0.5  &   2.2$\pm$0.1  &   2.4$\pm$0.2   &   1.0$\pm$0.1   &   0.4$\pm$0.1  &  0.3$\pm$0.1   &  \nodata       & \nodata        \\
G8.68           &  39.2 &  5.4  &   4.9$\pm$0.2  &   6.3$\pm$0.2  &   3.8$\pm$0.1  &   5.2$\pm$0.1   &   2.9$\pm$0.1   &   2.8$\pm$0.3  &  1.2$\pm$0.5   &   0.7$\pm$0.1  & \nodata        \\
G10.47          &  74.5 &  9.3  & 199.1$\pm$4.0  & 195.6$\pm$2.2  & 284.6$\pm$6.2  & 147.7$\pm$4.5   & 176.3$\pm$2.3   & 166.3$\pm$9.8  & 139.7$\pm$3.4  & 103.1$\pm$8.8  & 60.2$\pm$4.5   \\
G10.62          & -3.3  &  6.1  &  22.9$\pm$0.4  &  35.7$\pm$0.1  &  25.4$\pm$0.1  &  30.3$\pm$0.2   &  15.6$\pm$0.1   &  12.4$\pm$0.1  &   8.1$\pm$0.5  &  5.1$\pm$0.4   &  1.4$\pm$0.1   \\
I18182          &  59.1 &  6.6  &  12.4$\pm$2.0  &  20.7$\pm$2.2  &  12.7$\pm$0.4  &  17.4$\pm$0.8   &   9.0$\pm$0.7   &   8.2$\pm$0.7  &   8.3$\pm$0.8  &    \nodata     & \nodata        \\
G23.01          &  78.3 &  8.0  &  20.9$\pm$0.3  &  31.8$\pm$0.2  &  18.3$\pm$1.9  &  23.5$\pm$0.5   &  15.0$\pm$0.3   &  10.2$\pm$0.2  &  11.5$\pm$0.7  &  4.6$\pm$0.2   & \nodata        \\
G28.20N         &  95.3 &  5.0  &  38.9$\pm$1.6  &  40.0$\pm$1.6  &  28.5$\pm$0.7  &  33.3$\pm$0.4   &  20.9$\pm$0.2   &  19.4$\pm$1.0  &  12.3$\pm$0.5  &  4.8$\pm$0.6   & \nodata        \\
G31.41          &  99.6 &  7.6  &  65.8$\pm$1.3  &  63.3$\pm$7.6  &  71.3$\pm$1.2  &  69.8$\pm$1.0   &  58.8$\pm$0.6   &  47.5$\pm$3.1  &  35.4$\pm$0.7  & 20.8$\pm$1.4   & 12.4$\pm$0.2   \\
I18566          &  84.6 &  8.0  &  13.5$\pm$1.0  &  11.9$\pm$0.5  &  13.5$\pm$0.2  &  13.8$\pm$0.2   &   8.0$\pm$0.3   &   7.4$\pm$0.2  &   4.9$\pm$0.9  &   \nodata      & \nodata        \\
G45.07          &  58.5 &  7.4  &  47.6$\pm$5.9  &  52.4$\pm$0.7  &  38.3$\pm$0.4  &  47.0$\pm$0.3   &  20.4$\pm$0.2   &  19.2$\pm$1.0  &  13.3$\pm$1.0  &  5.2$\pm$0.8   & \nodata        \\
G45.47          &  64.3 &  4.2  &   9.5$\pm$0.9  &   9.3$\pm$0.9  &   7.0$\pm$0.3  &   7.7$\pm$0.2   &   2.4$\pm$0.3   &   1.3$\pm$0.2  &   1.1$\pm$0.2  &  1.1$\pm$0.2   & \nodata        \\
W51e2           &  55.7 &  8.0  & 252.0$\pm$9.8  & 408.8$\pm$9.5  & 413.8$\pm$5.5  & 449.2$\pm$5.1   & 270.2$\pm$5.2   & 235.4$\pm$9.1  & 236.6$\pm$9.3  & 89.5$\pm$3.2   & 47.2$\pm$2.9   \\
W51e8           &  58.5 &  9.0  & 191.3$\pm$3.3  & 318.0$\pm$9.4  & 311.1$\pm$2.2  & 311.8$\pm$2.2   & 191.8$\pm$3.9   & 109.5$\pm$9.9  & 191.9$\pm$6.7  & 91.5$\pm$3.6   & 27.7$\pm$7.8   
\enddata
\tablecomments{(a) Average values from all detected $K$-components}
\end{deluxetable*}

\newpage

\begin{deluxetable*}{lcrlccccccc}
\tabletypesize{\scriptsize}
\tablewidth{0pc}
\tablecaption{Physical parameters derived from \chtcnte \label{xclassrd}}
\tablehead{
\colhead{}& \multicolumn{5}{c}{XCLASS Program\tablenotemark{a}} &\colhead{}& \multicolumn{2}{c}{Rotational Diagram\tablenotemark{b}}& & \\ \cline{2-6}  \cline{8-9}  
\colhead{}&\colhead{$\theta_{\rm s}$}&\colhead{T$_{\rm rot}$\tablenotemark{c}}&\colhead{$N_{\rm tot}$\tablenotemark{c}}&\colhead{$\Delta v$}&\colhead{$X_{\rm CH3CN}^{\rm core}$}&\colhead{}&\colhead{T$_{\rm rot}$}&\colhead{$N_{\rm tot}$}&\colhead{}&\mv \tablenotemark{d} \\  
\colhead{Source}&\colhead{(arcsec)}&\colhead{(K)}&\colhead{(cm$^{-2}$)}&\colhead{(km/s)}& \colhead{}&\colhead{}&\colhead{(K)}&\colhead{(cm$^{-2}$)}&\colhead{}& (\mo) }  
 \startdata                                                   
W3OH            & 1.1              & 122$^{+56}_{-12}$  & 3.3$^{+2.1}_{-3.5}$(15) & 5.0            & 1.9(-9)  &          &  90         & 2.2(14) &  &   58  \\ 
                & 5.5              & 68$^{+22}_{-25}$   & 7.5$^{+1.1}_{-2.0}$(13) & 6.0            &          &          &             &          &  &       \\ 
W3TW            & 0.8              & 367$^{+92}_{-102}$ & 3.9$^{+0.6}_{-1.1}$(16) & 7.0            & 3.2(-8)  &          &  182        & 6.9(14) &  &   85  \\ 
                & 3.5              & 108$^{+43}_{-6}$   & 7.1$^{+2.5}_{-0.6}$(14) & 8.0            &          &          &             &          &  &       \\ 
I16547          & 0.7              & 272$^{+93}_{-59}$  & 2.2$^{+0.4}_{-0.6}$(16) & 7.0            & 4.5(-8)  &          &  245        & 2.0(15) &  &   235 \\ 
                & 2.4              & 78$^{+20}_{-38}$   & 8.8$^{+2.2}_{-1.3}$(13) & 8.0            &          &          &             &          &  &       \\ 
I17233          & 0.9              & 346$^{+137}_{-105}$& 2.4$^{+1.9}_{-3.2}$(17) & 9.0            & 1.8(-7)  &          &  408        & 2.0(16) &  &   138 \\ 
                & 6.3              & 132$^{+97}_{-100}$ & 9.1$^{+1.9}_{-0.9}$(13) & 9.0            &          &          &             &          &  &       \\ 
G5.89           & 0.9              & 165$^{+70}_{-33}$  & 9.6$^{+1.0}_{-2.2}$(14) & 5.0            & 3.6(-9)  &          &  104        & 2.5(13) &  &   74  \\ 
                & 7.0              & 40$^{+20}_{-15}$   & 1.6$^{+0.2}_{-3.1}$(13) & 7.0            &          &          &             &          &  &       \\ 
G8.68           & 1.0              & 281$^{+92}_{-79}$  & 4.2$^{+0.9}_{-1.2}$(15) & 5.0            & 4.2(-8)  &          &  202        & 9.5(13) &  &   209 \\ 
                & 4.7              & 77$^{+28}_{-30}$   & 1.9$^{+0.6}_{-0.9}$(14) & 6.0            &          &          &             &          &  &       \\ 
G10.47          & 0.6              & 408$^{+96}_{-108}$ & 5.1$^{+2.4}_{-1.9}$(17) & 6.0            & 6.1(-8)  &          &  499        & 2.5(16) &  &   400 \\ 
                & 4.0              & 82$^{+20}_{-22}$   & 4.1$^{+0.9}_{-0.6}$(14) & 7.0            &          &          &             &          &  &       \\ 
G10.62          & 1.9              & 415$^{+55}_{-123}$ & 6.7$^{+0.7}_{-0.4}$(15) & 6.0            & 1.0(-8)  &          &  224        & 6.5(14) &  &   310 \\ 
                & 6.6              & 95$^{+25}_{-45}$   & 3.0$^{+2.0}_{-0.8}$(14) & 8.0            &          &          &             &          &  &       \\ 
I18182          & 1.1              & 219$^{+102}_{-110}$& 7.3$^{+2.6}_{-1.8}$(15) & 5.0            & 2.1(-8)  &          &  256        & 4.3(14) &  &   213 \\ 
                & 4.4              & 75$^{+45}_{-50}$   & 8.2$^{+2.2}_{-1.3}$(13) & 6.0            &          &          &             &          &  &       \\ 
G23.01          & 0.4              & 237$^{+93}_{-39}$  & 1.5$^{+0.7}_{-0.7}$(17) & 7.0            & 1.0(-7)  &          &  231        & 6.6(14) &  &   407 \\ 
                & 4.6              & 58$^{+92}_{-28}$   & 1.7$^{+1.1}_{-0.5}$(14) & 8.0            &          &          &             &          &  &       \\ 
G28.20N         & 0.6              & 295$^{+98}_{-92}$  & 6.2$^{+1.1}_{-0.9}$(16) & 5.0            & 8.0(-8)  &          &  302        & 1.1(15) &  &   100 \\ 
                & 5.0              & 59$^{+31}_{-30}$   & 2.4$^{+1.5}_{-1.8}$(14) & 6.0            &          &          &             &          &  &       \\ 
G31.41          & 0.4              & 327$^{+173}_{-97}$ & 1.5$^{+1.3}_{-0.5}$(17) & 5.0            & 4.2(-8)  &          &  402        & 4.8(15) &  &   300 \\ 
                & 4.6              & 95$^{+75}_{-55}$   & 9.1$^{+5.9}_{-5.0}$(13) & 7.0            &          &          &             &          &  &       \\ 
I18566          & 1.1              & 382$^{+107}_{-90}$ & 7.1$^{+1.2}_{-2.1}$(15) & 8.0            & 6.4(-8)  &          &  308        & 4.9(14) &  &   473 \\ 
                & 4.2              & 110$^{+62}_{-60}$  & 5.2$^{+1.0}_{-1.0}$(13) & 9.0            &          &          &             &          &  &       \\ 
G45.07          & 1.1              & 290$^{+89}_{-65}$  & 5.4$^{+0.9}_{-2.1}$(15) & 6.0            & 2.0(-9)  &          &  200        & 8.5(14) &  &   270 \\ 
                & 3.7              & 82$^{+25}_{-39}$   & 4.1$^{+1.1}_{-1.8}$(14) & 7.0            &          &          &             &          &  &       \\ 
G45.47          & 1.2              & 155$^{+95}_{-60}$  & 6.8$^{+1.8}_{-1.3}$(14) & 5.0            & 1.0(-9)  &          &  131        & 8.0(13) &  &   108 \\ 
                & 4.0              & 65$^{+18}_{-23}$   & 3.8$^{+1.2}_{-0.7}$(13) & 5.0            &          &          &             &          &  &       \\ 
W51e2           & 0.5              & 485$^{+121}_{-130}$& 2.8$^{+2.3}_{-1.9}$(17) & 6.0            & 7.0(-8)  &          &  314        & 1.6(16) &  &   194 \\ 
                & 3.0              & 118$^{+44}_{-42}$  & 1.0$^{+1.4}_{-1.7}$(15) & 8.0            &          &          &             &          &  &       \\ 
W51e8           & 0.5              & 384$^{+126}_{-150}$& 1.8$^{+0.8}_{-0.5}$(17) & 7.0            & 3.4(-8)  &          &  304        & 1.2(16) &  &   203 \\ 
                & 3.2              & 85$^{+95}_{-35}$   & 1.8$^{+0.3}_{-0.4}$(14) & 9.0            &          &          &             &          &  &     
\enddata
\tablecomments{(a) Final values of source sizes, rotational temperatures, column densities, and line widths for the best fit 
of the synthetic to the observed spectrum. All sources fit with two components. (b) Rotational temperatures and column densities 
from the best linear fit, in which we included the $K$-lines 0 and 1. (c) Uncertainties in temperature and column density from 
XCLASS were estimated with synthetic spectra, perturbing both parameters and considering over/undershoot by 20\% the $K=2$ transition 
(see Section \ref{xclass}). (d) Virial mass was estimated using $\Delta V$ from Table \ref{linedata}.} 
\end{deluxetable*}

\newpage

\begin{deluxetable*}{lccc}
\tabletypesize{\scriptsize}
\tablewidth{0pc} 
\tablecaption{Velocity-integrated emission (moment 0) of \chtcnte \label{velint}} 
\tablehead{
\colhead{}      & \colhead{}        & \colhead{\chtcn~Transitions} &  \colhead{}                  \\   \cline{2-4} 
\colhead{}      & \colhead{$K=3$}   & \colhead{$K=5$}              & \colhead{$K=7$\tablenotemark{a}}    \\         
\colhead{Sources}& \colhead{(\jpbpv)}& \colhead{(\jpbpv)}           & \colhead{(\jpbpv)} }
 \startdata   
W3OH and TW     & 29.1  & 17.4   & 3.9    \\          
I16547          & 6.2   & 4.8    & 2.3    \\          
I17233          & 224.7 & 206.6  & 78.6   \\          
G5.89           & 5.8   & 2.5    & 1.8\tablenotemark{a} \\          
G8.68           & 7.6   & 4.2    & 1.4    \\          
G10.47          & 63.1  & 100.6  & 45.4   \\          
G10.62          & 33.6  & 19.1   & 6.2    \\          
I18182          & 11.8  & 6.5    & 8.0\tablenotemark{a}  \\          
G23.01          & 18.7  & 11.3   & 3.5    \\          
G28.20N         & 13.9  & 8.9    & 2.5    \\          
G31.41          & 32.8  & 31.1   & 11.8   \\          
I18566          & 7.3   & 4.7    & 1.3    \\          
G45.07          & 10.8  & 4.2    & 1.6    \\          
G45.47          & 3.5   & 0.6    & 0.9    \\          
W51e2 and W51e8 & 34.6  & 26.6   & 13.0    
\enddata
\tablecomments{(a) For G5.89--0.37 and I18182 we present the line $K=6$.}
\end{deluxetable*}

\newpage

\begin{figure*}
 \epsfig{file=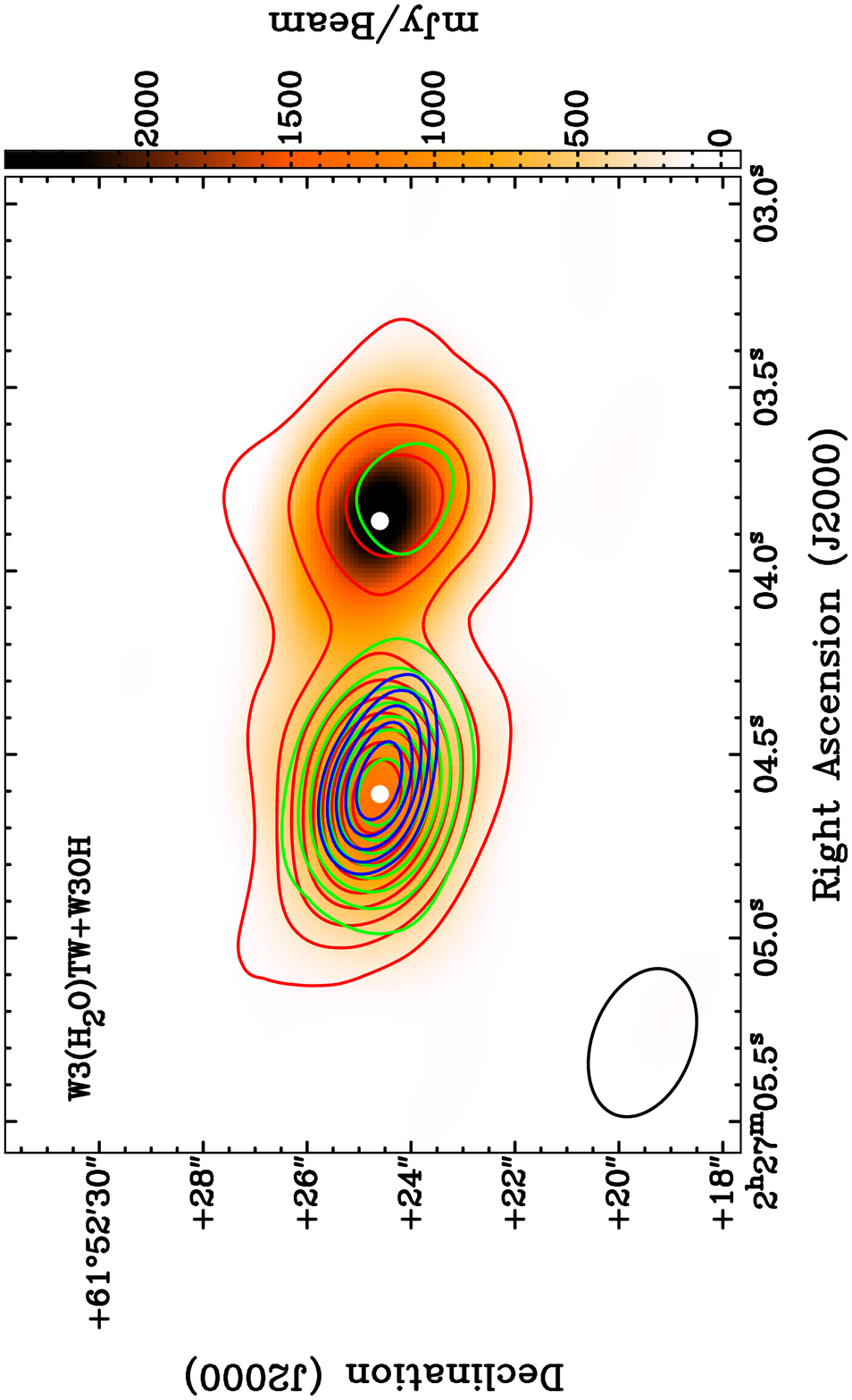, angle=-90, width=0.5\linewidth}
 \epsfig{file=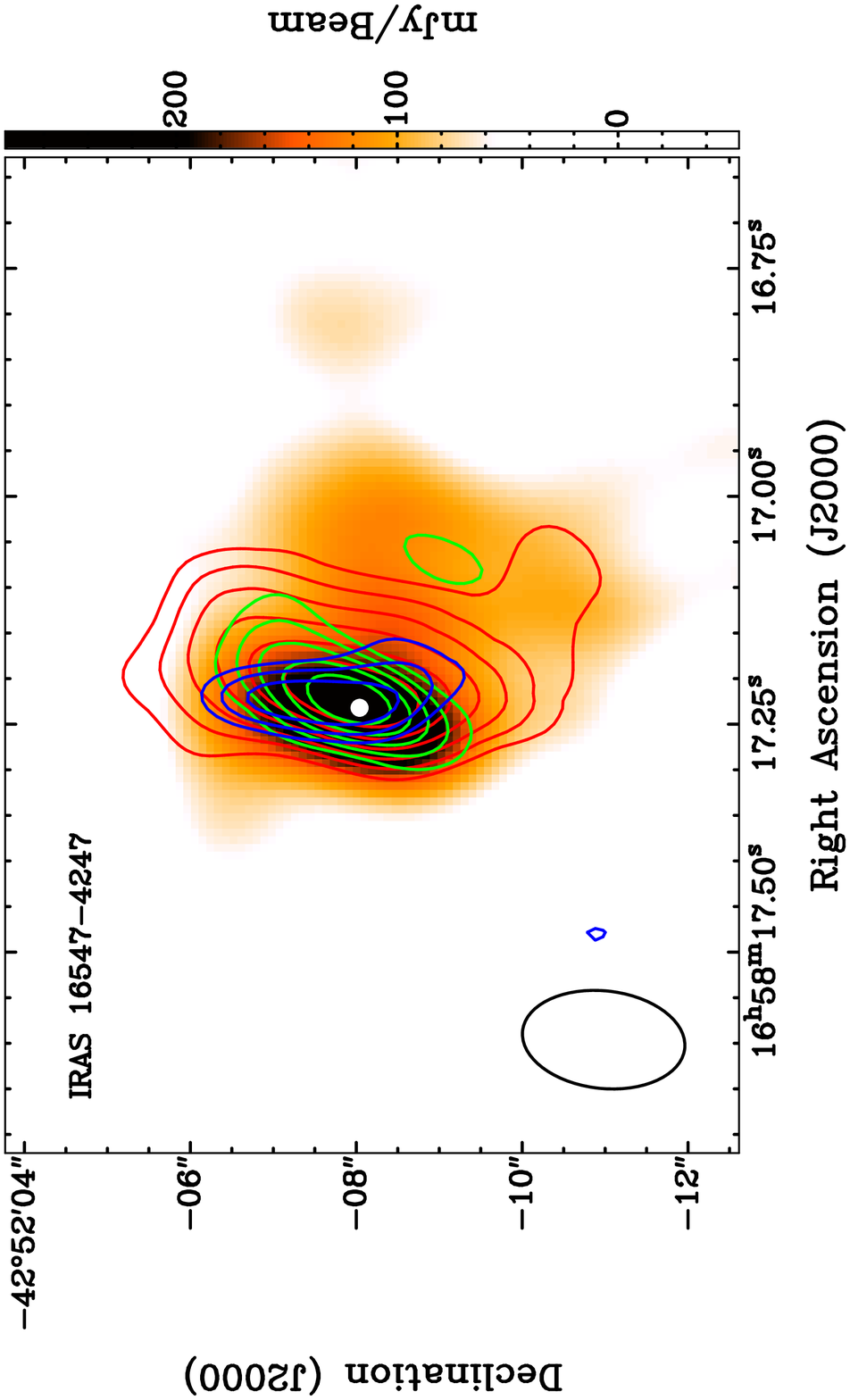, angle=-90, width=0.5\linewidth}
 \epsfig{file=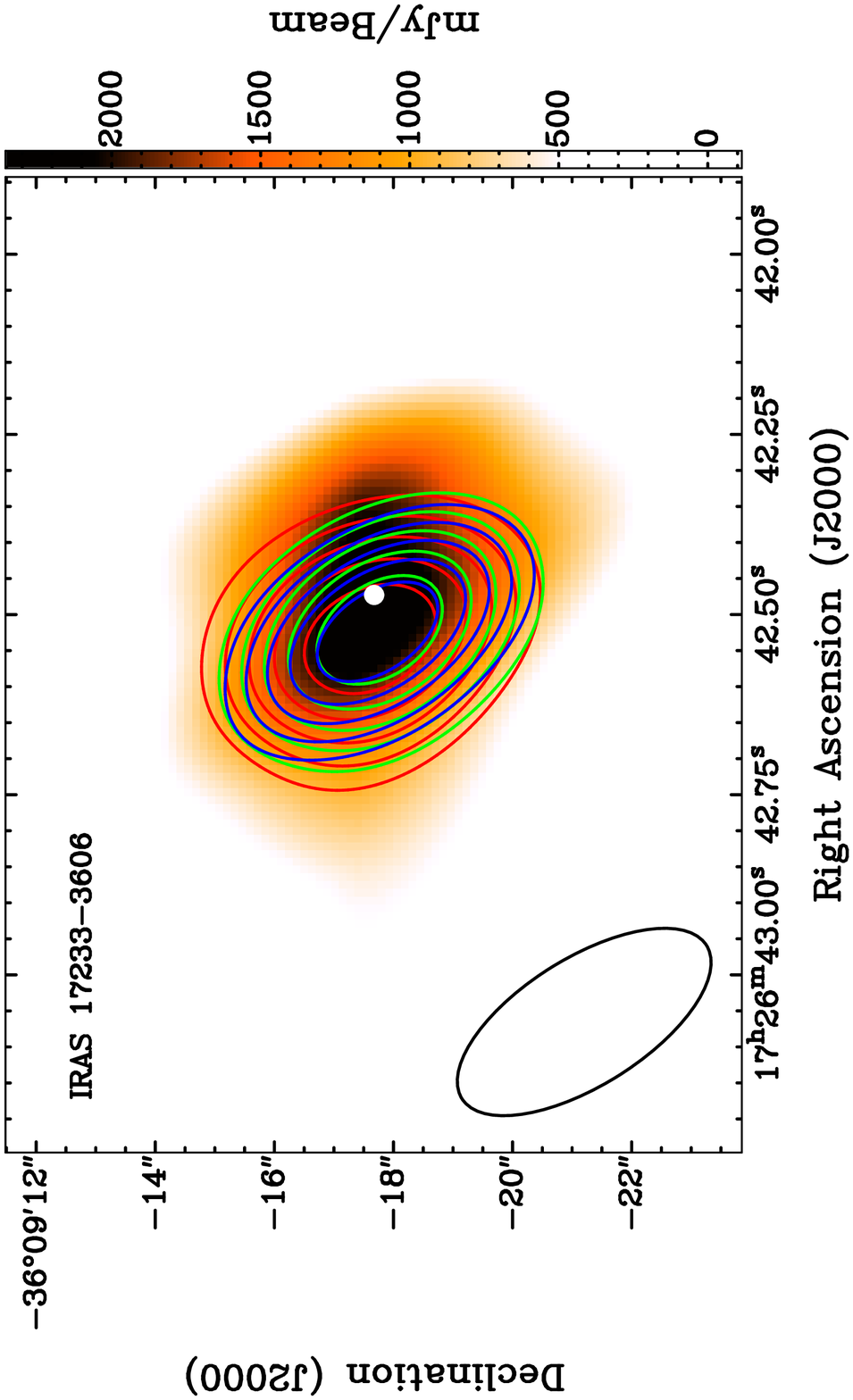, angle=-90, width=0.5\linewidth}
 \epsfig{file=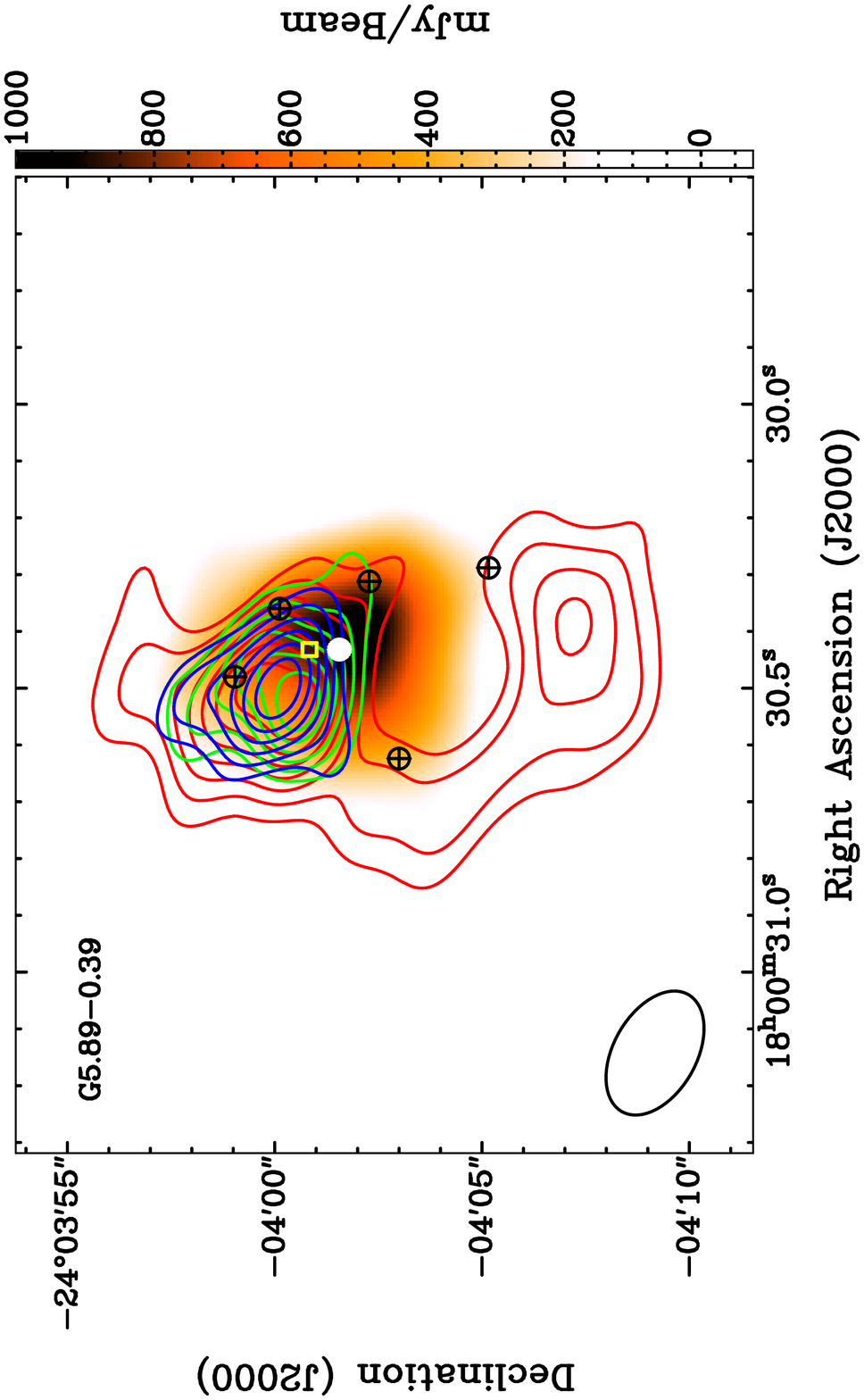, angle=-90, width=0.5\linewidth}
 \epsfig{file=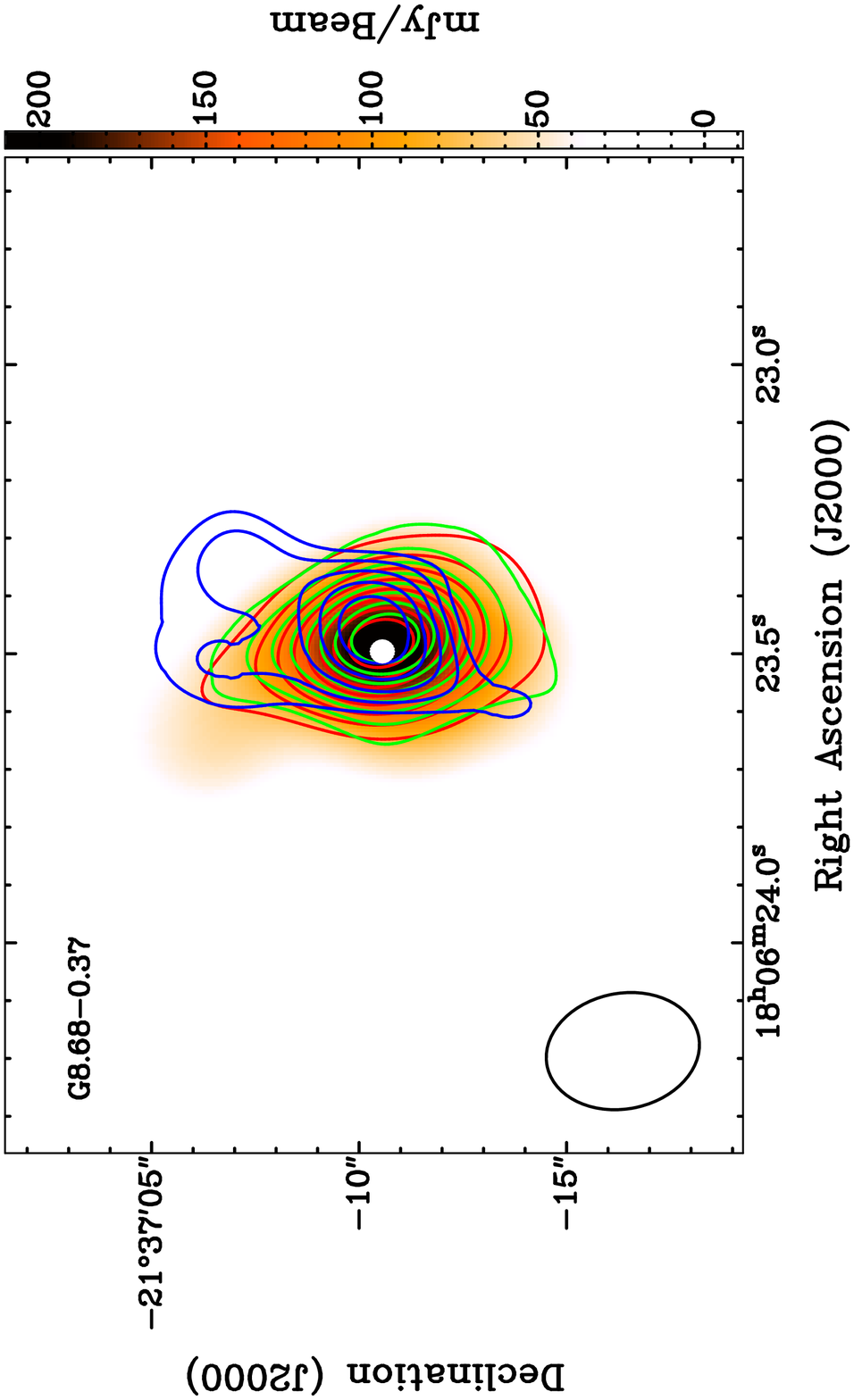, angle=-90, width=0.5\linewidth}
 \epsfig{file=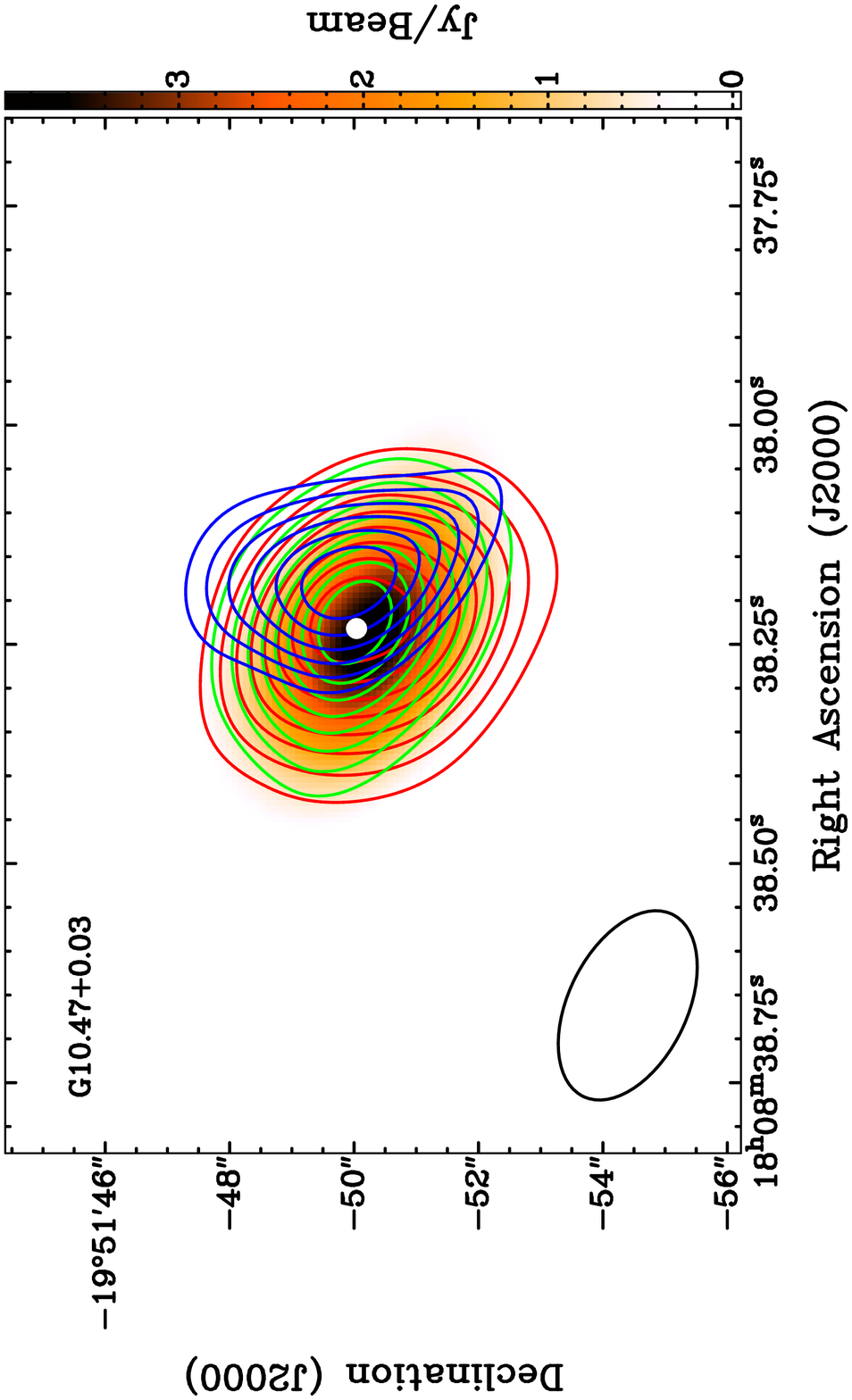, angle=-90, width=0.5\linewidth}
\caption{\scriptsize SMA line-free continuum maps at 1.3\,mm (color scale) and velocity-integrated emission (moment 0, contours) 
of \chtcnte~for $K=3$ (red), $K=5$ (green) and $K=7$ (blue) lines.  
In all cases the contour levels have steps of 10\% until 90\% of 
the integrated emission shown in Table 6.
For W3OH-H2O contour levels begin at 10\% ($K=3$), 20\% ($K=5$), and 50\% ($K=7$). 
For I16547 contour levels begin at 40\% ($K=3$), 50\% ($K=5$), and 70\% ($K=7$).  
For I17233 contour levels begin at 50\% ($K=3$, $K=5$, and $K=7$).
For G5.89 we used the $K$--lines 3, 5, and 6 with contour levels beginning at 30\%, 40\%, and 40\%,  respectively.
For G5.89 the yellow box marks the position 
of the Feldt's star (Feldt et al. 2003), and circles with crosses show condensations with excess 
870 $\mu$m emission reported by Hunter et al. (2008).
For G8.68 contour levels begin at 20\% ($K=3$ and $K=5$) and 50\% ($K=7$). 
For G10.47 contour levels begin at 20\% ($K=3$ and $K=5$) and 40\% ($K=7$).
In all cases a white dot marks the peak position of 1.3\,mm continuum emission (Table \ref{mmcont}), and the 
synthesized beam (Table \ref{mmcont}) is shown at the bottom-left.\label{lines1} }
\end{figure*}

\newpage

\begin{figure*} 
 \epsfig{file=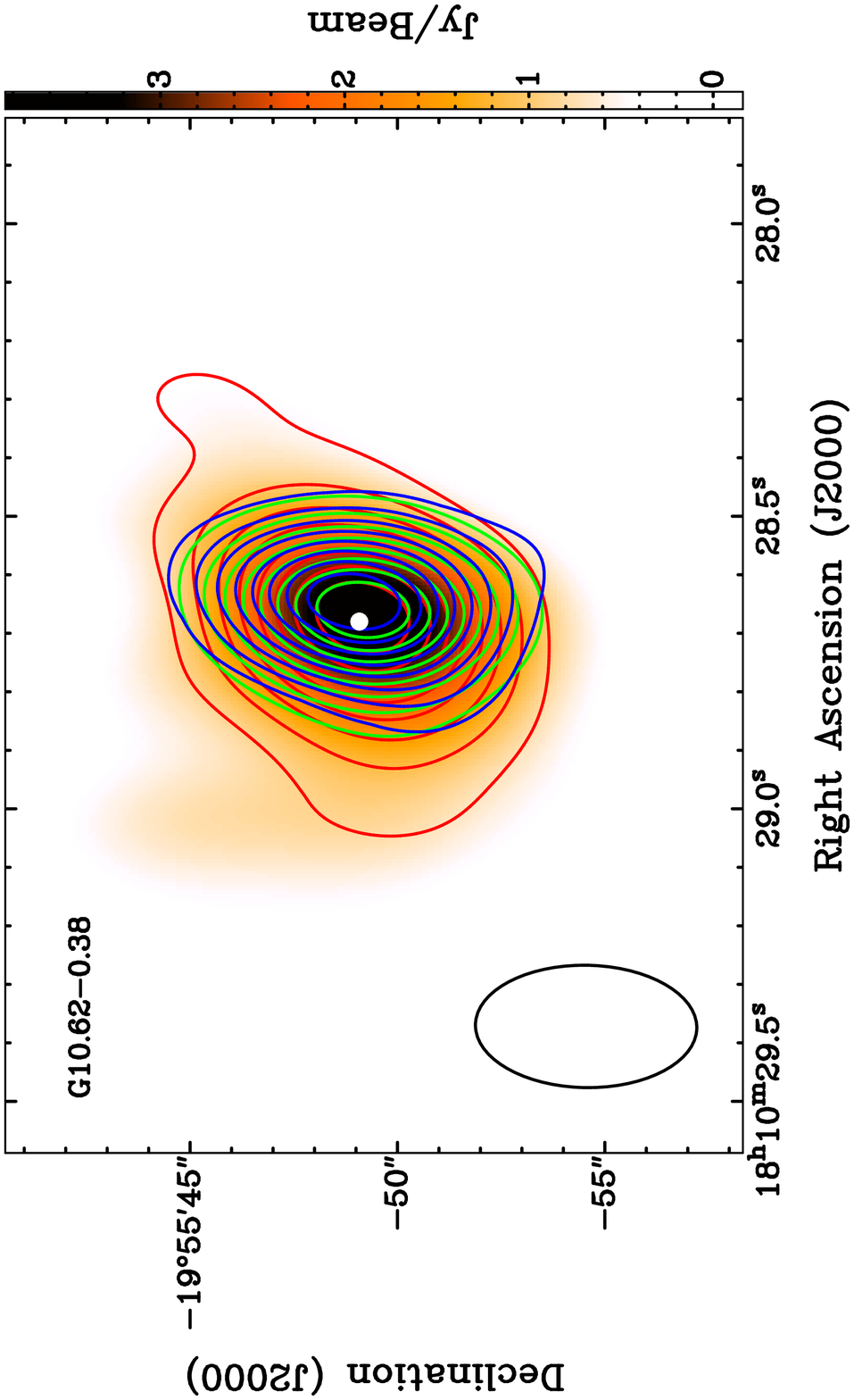, angle=-90, width=0.5\linewidth}
 \epsfig{file=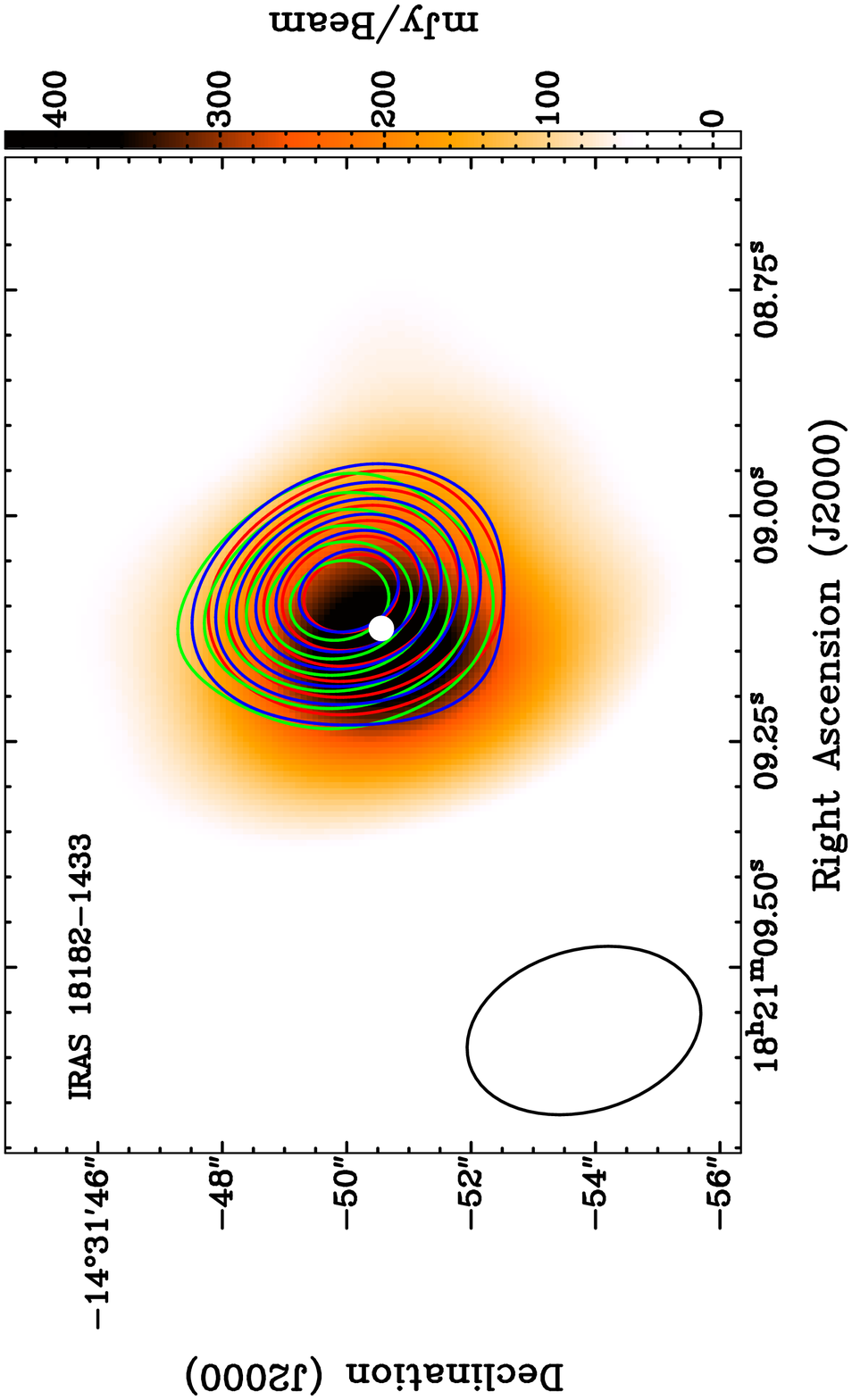, angle=-90, width=0.5\linewidth} 
 \epsfig{file=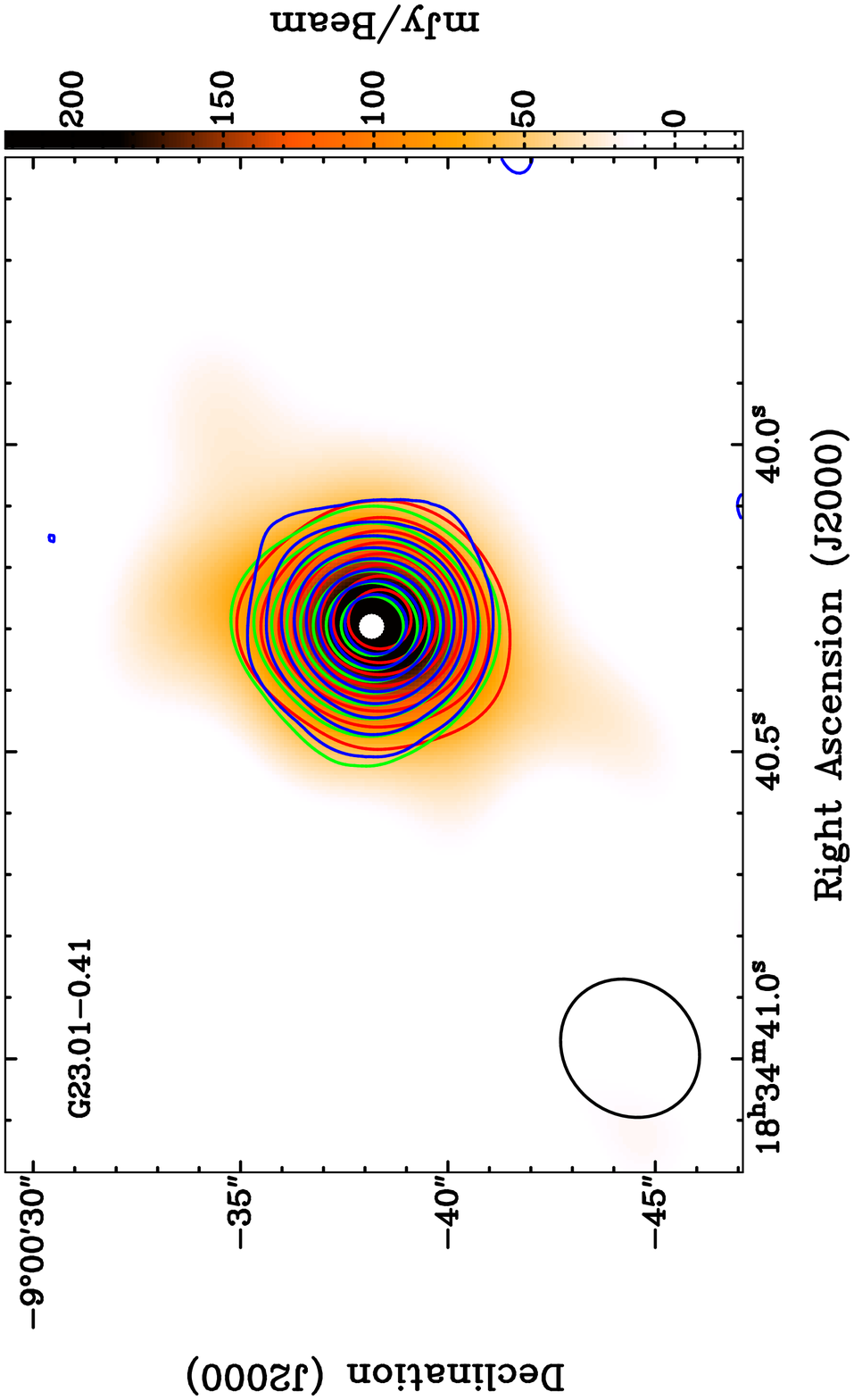, angle=-90, width=0.5\linewidth}
 \epsfig{file=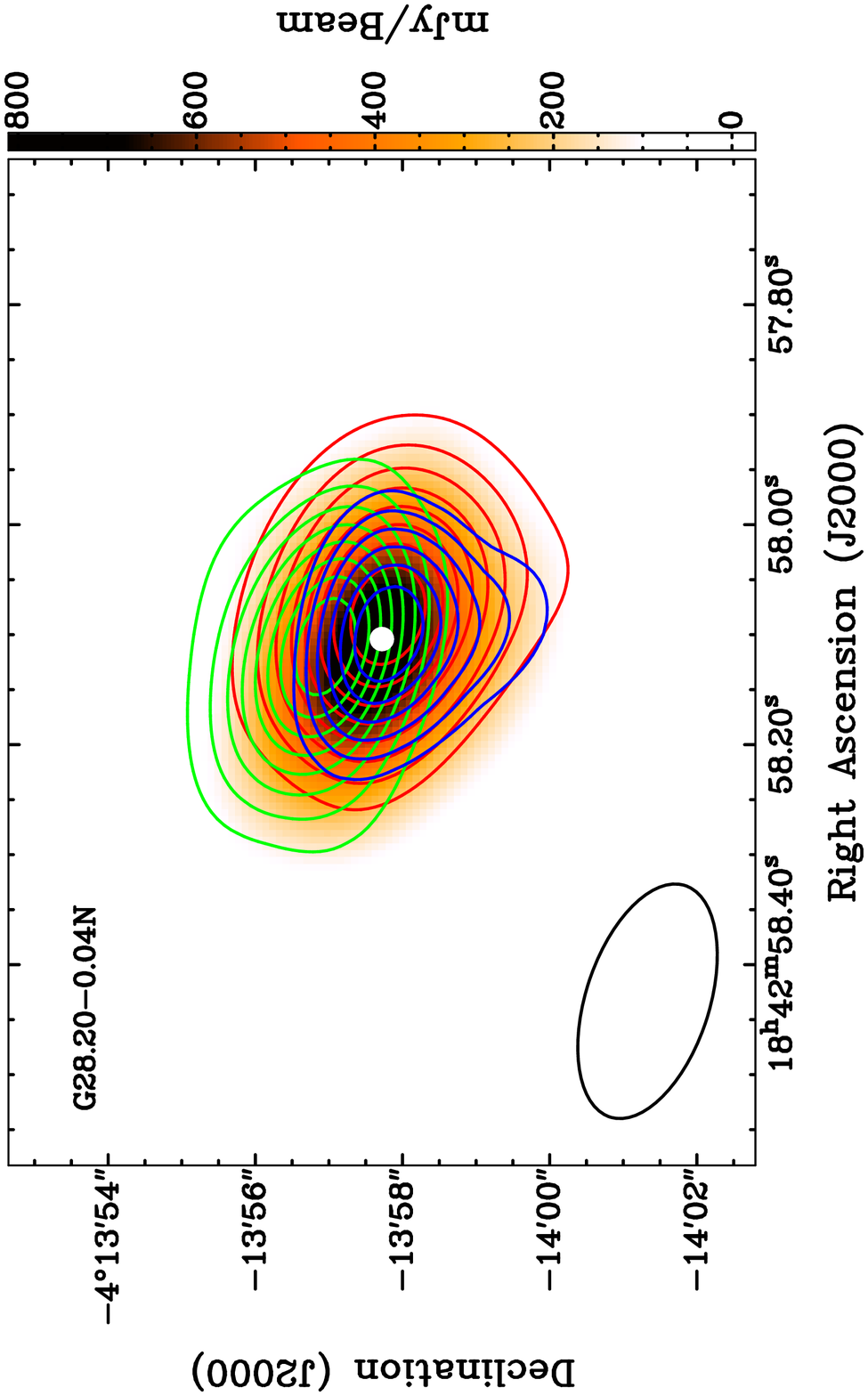, angle=-90, width=0.5\linewidth}
 \epsfig{file=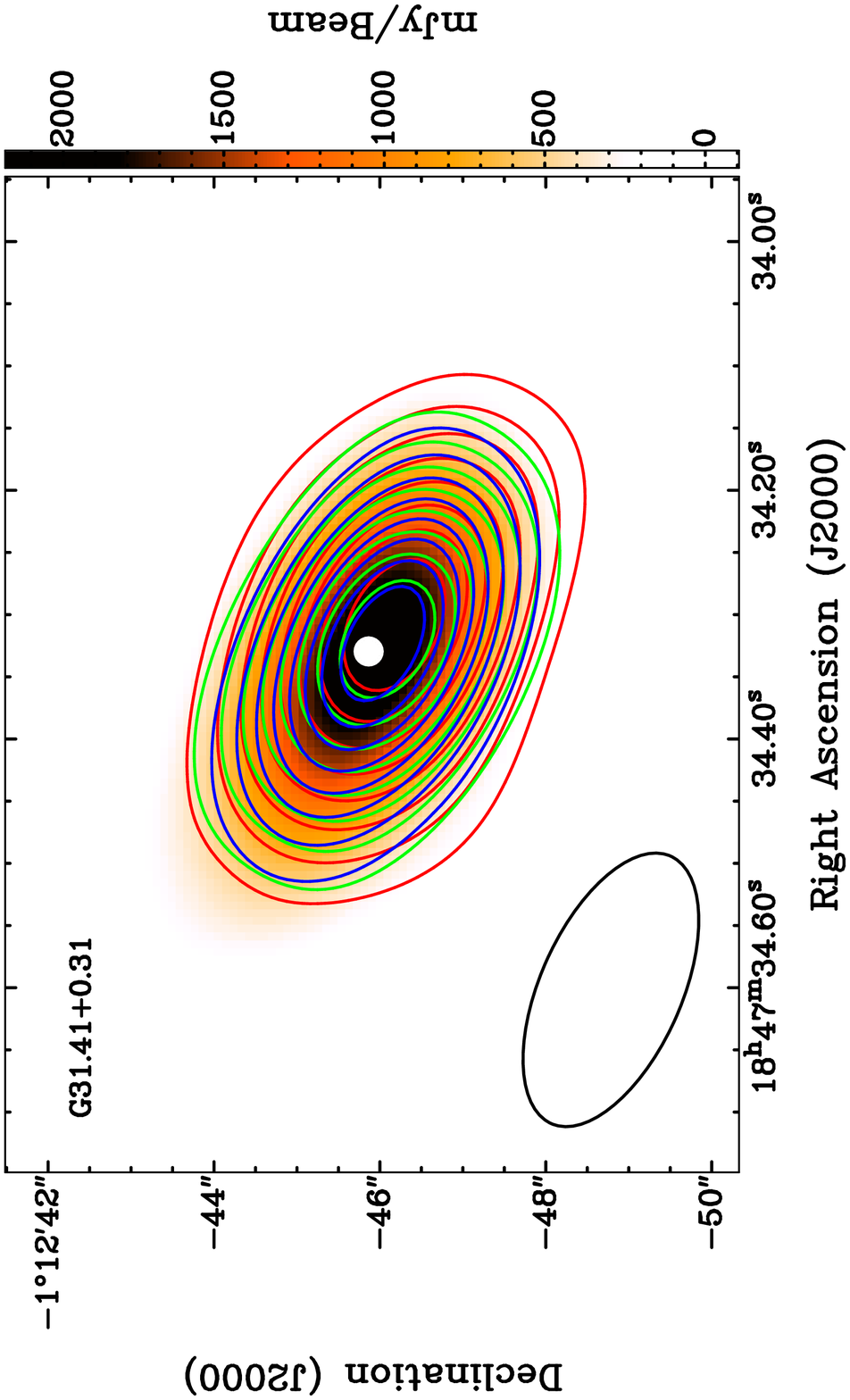, angle=-90, width=0.5\linewidth}
 \epsfig{file=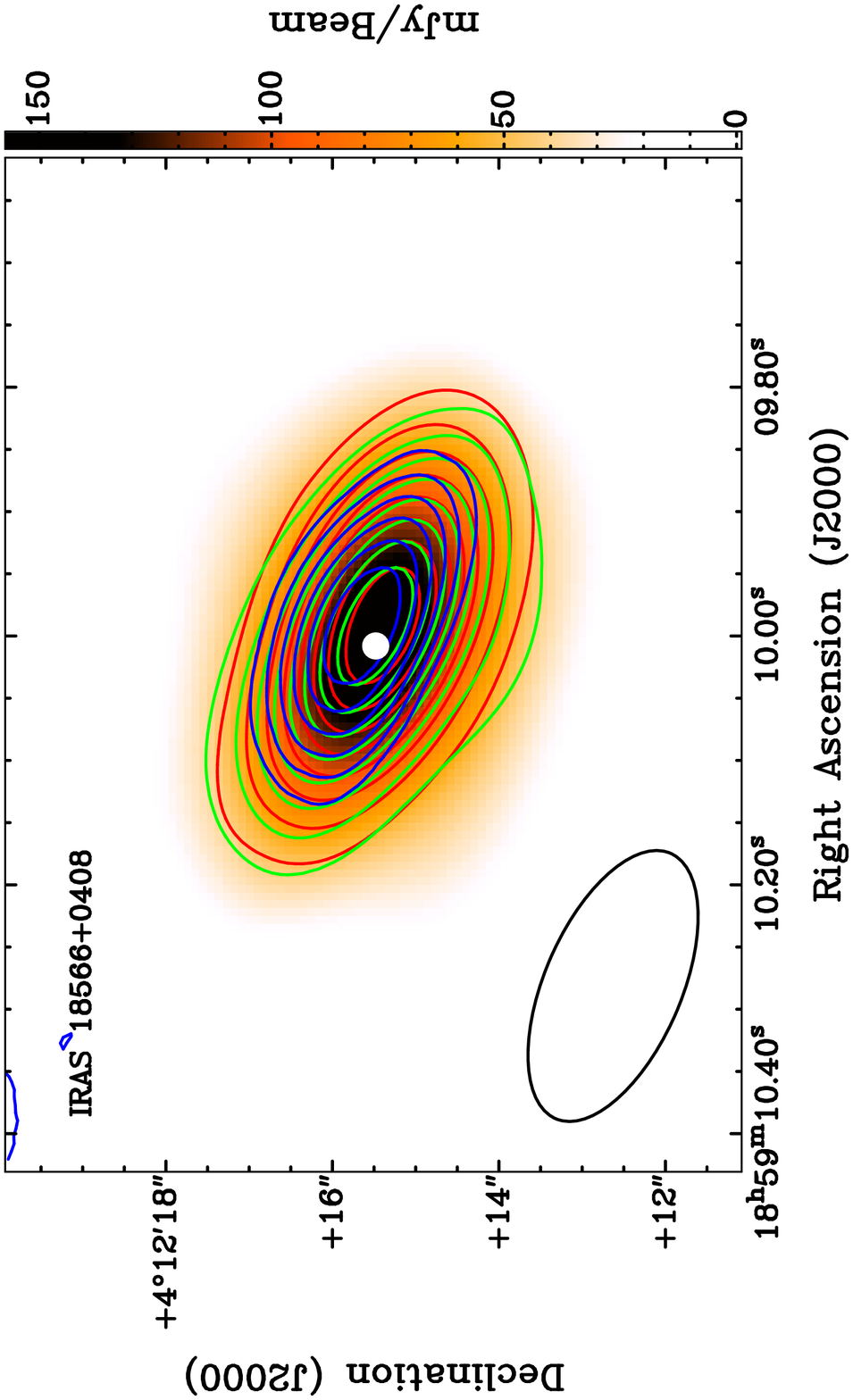, angle=-90, width=0.5\linewidth}
 \caption{Continuation of Figure \ref{lines1}.
For G10.62 contour levels begin at 20\% ($K=3$, $K=5$, and $K=7$).
For I18182 we used the $K$--lines 3, 5 and 6 with contour levels beginning at 50\%.
For G23.01 contour levels begin at 20\% ($K=3$, $K=5$, and $K=7$).
For G28.20 contour levels begin at 20\% ($K=3$ and $K=5$), and 40\% ($K=7$). 
For G31.41 contour levels begin at 40\% ($K=3$, $K=5$, and $K=7$).
For I18566 contour levels begin at 20\% ($K=3$ and $K=5$), and 40\% ($K=7$).\label{lines2}}
\end{figure*} 

\newpage

\begin{figure*}
 \epsfig{file=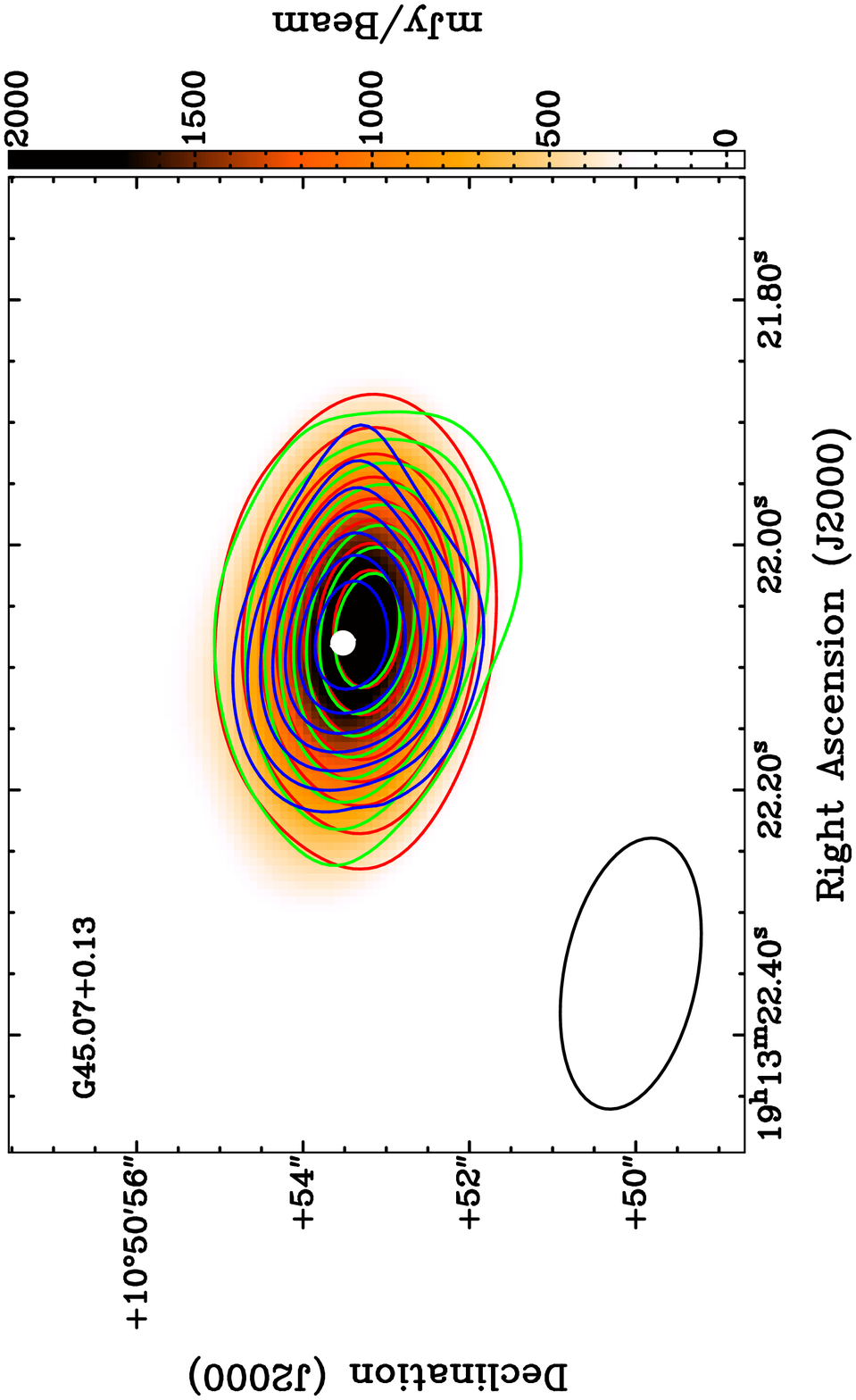, angle=-90, width=0.5\linewidth}
 \epsfig{file=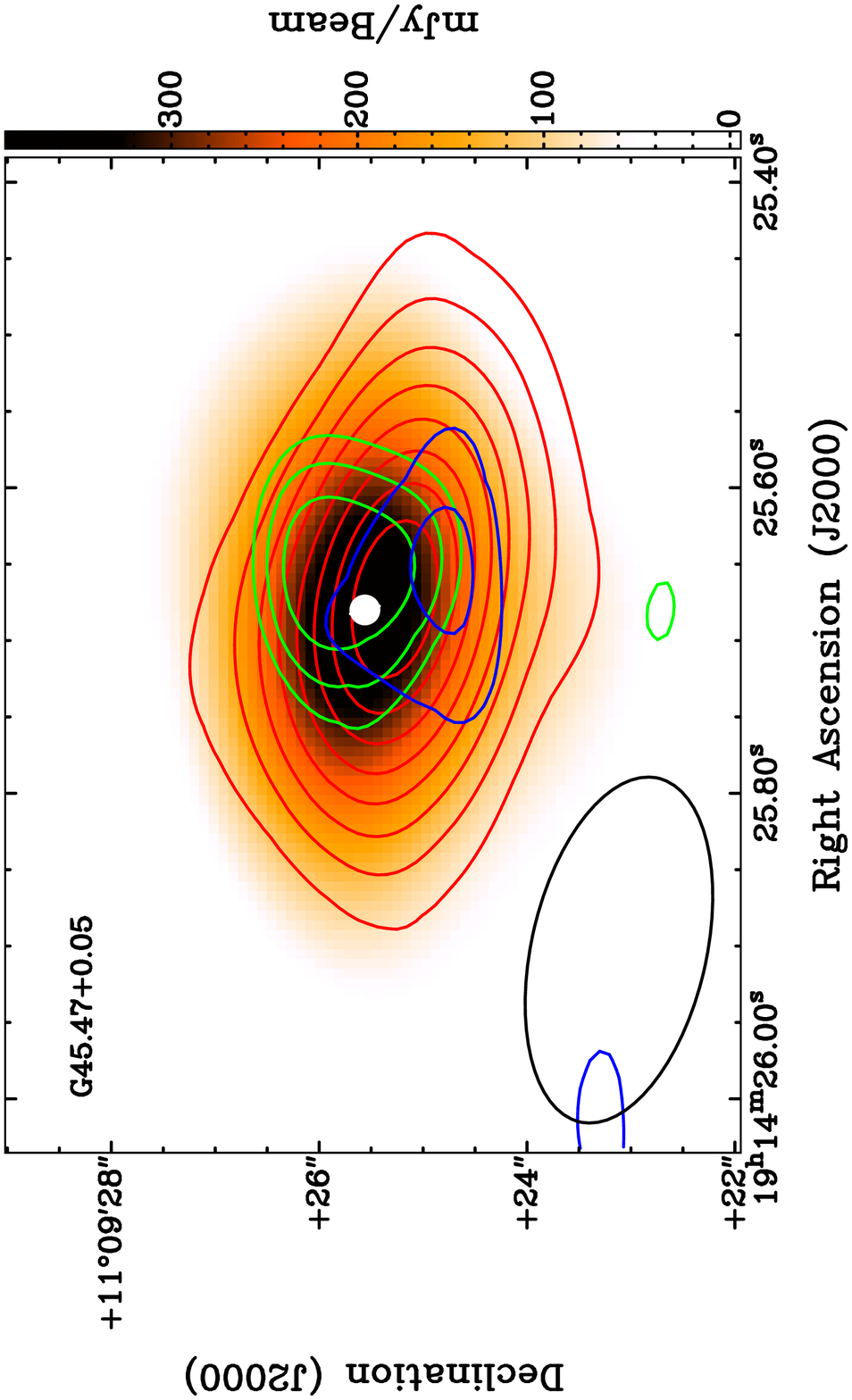, angle=-90, width=0.5\linewidth}
 \epsfig{file=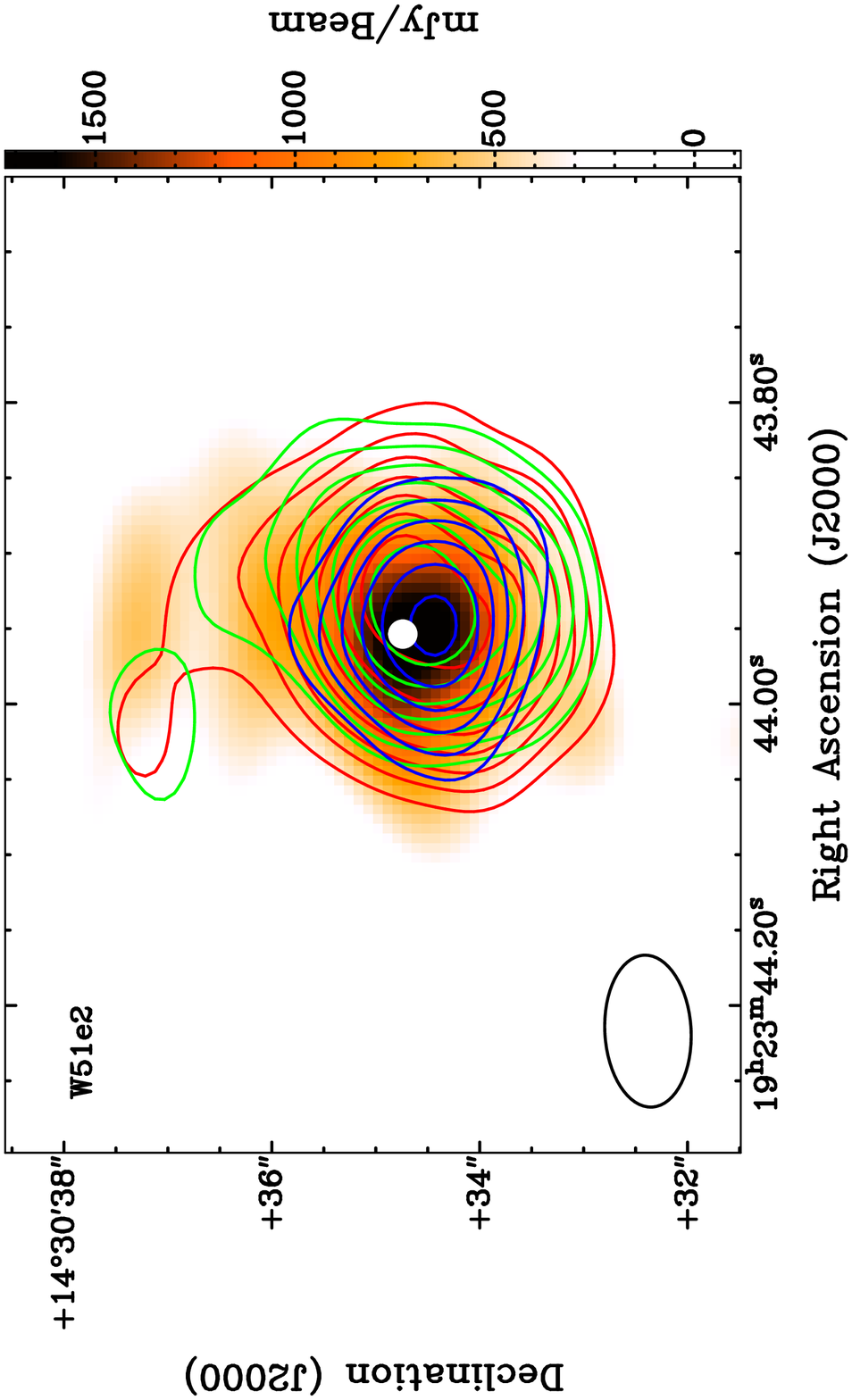, angle=-90, width=0.5\linewidth}
 \epsfig{file=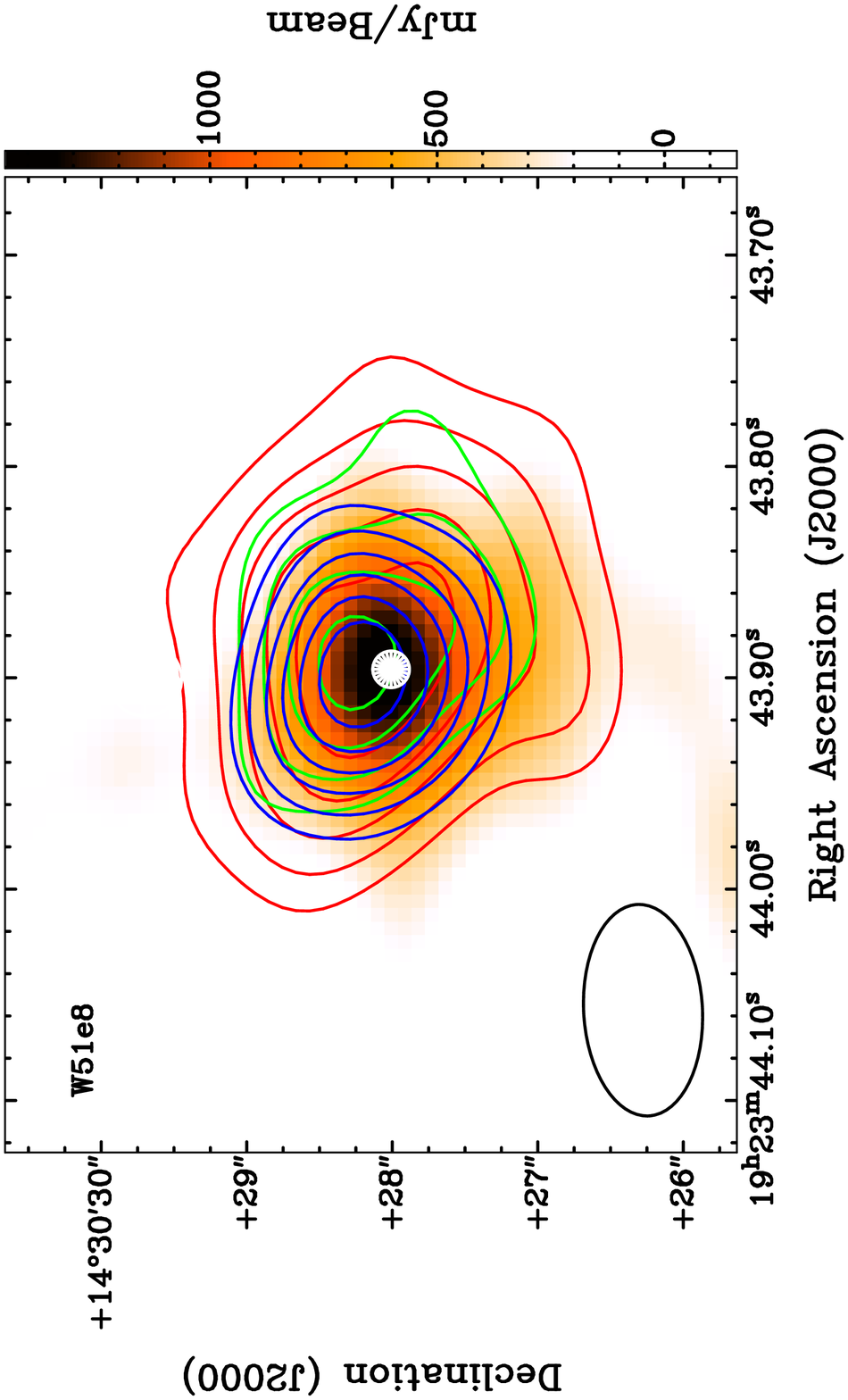, angle=-90, width=0.5\linewidth} 
 \caption{Continuation of Figure \ref{lines1}. 
For G45.07 contour levels begin at 20\% ($K=3$ and $K=5$), and 30\% ($K=7$). 
For G45.47 contour levels begin at 20\% ($K=3$), 70\% ($K=5$), and 50\% ($K=7$).
For W51e2 contour levels begin at 30\%  ($K=3$ and $K=5$), and 40\% ($K=7$). 
For W51e8 contour levels begin at 30\%  ($K=3$), and 40\% ($K=5$ and $K=7$).\label{lines3}}
\end{figure*}

\newpage
 
\begin{figure*} 
\centering
\includegraphics[angle=0,width=0.95\textwidth]{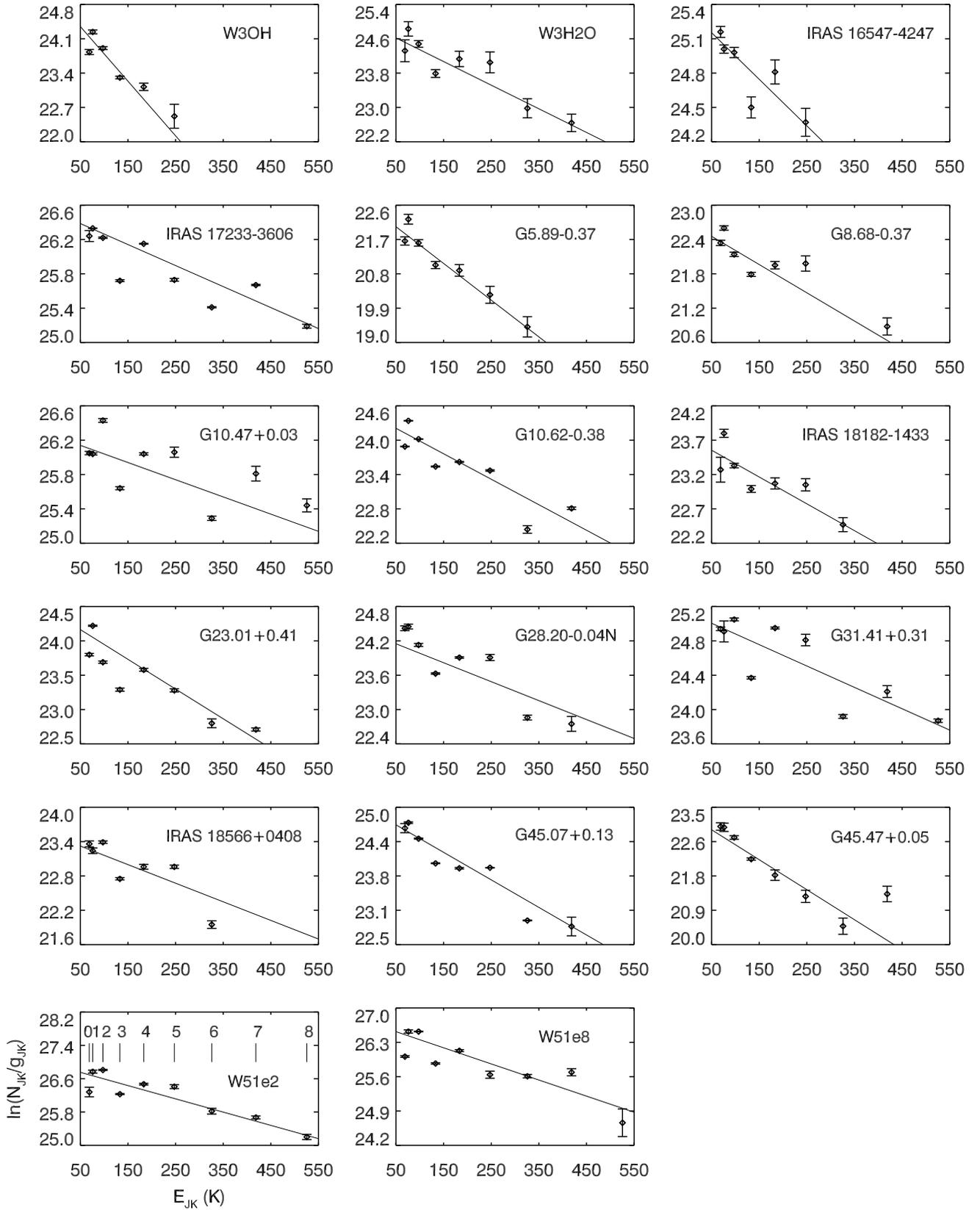}
\caption{Rotation diagrams for \chtcnte. The line is the linear fit of all data points 
           in the plots. The numbers in the lower left panel, of W51e2, represent the $K-$ladder 
           quantum numbers.\label{RDs}}
\end{figure*}

\newpage

\begin{figure*} 
\centering
\includegraphics[angle=0,width=0.95\textwidth]{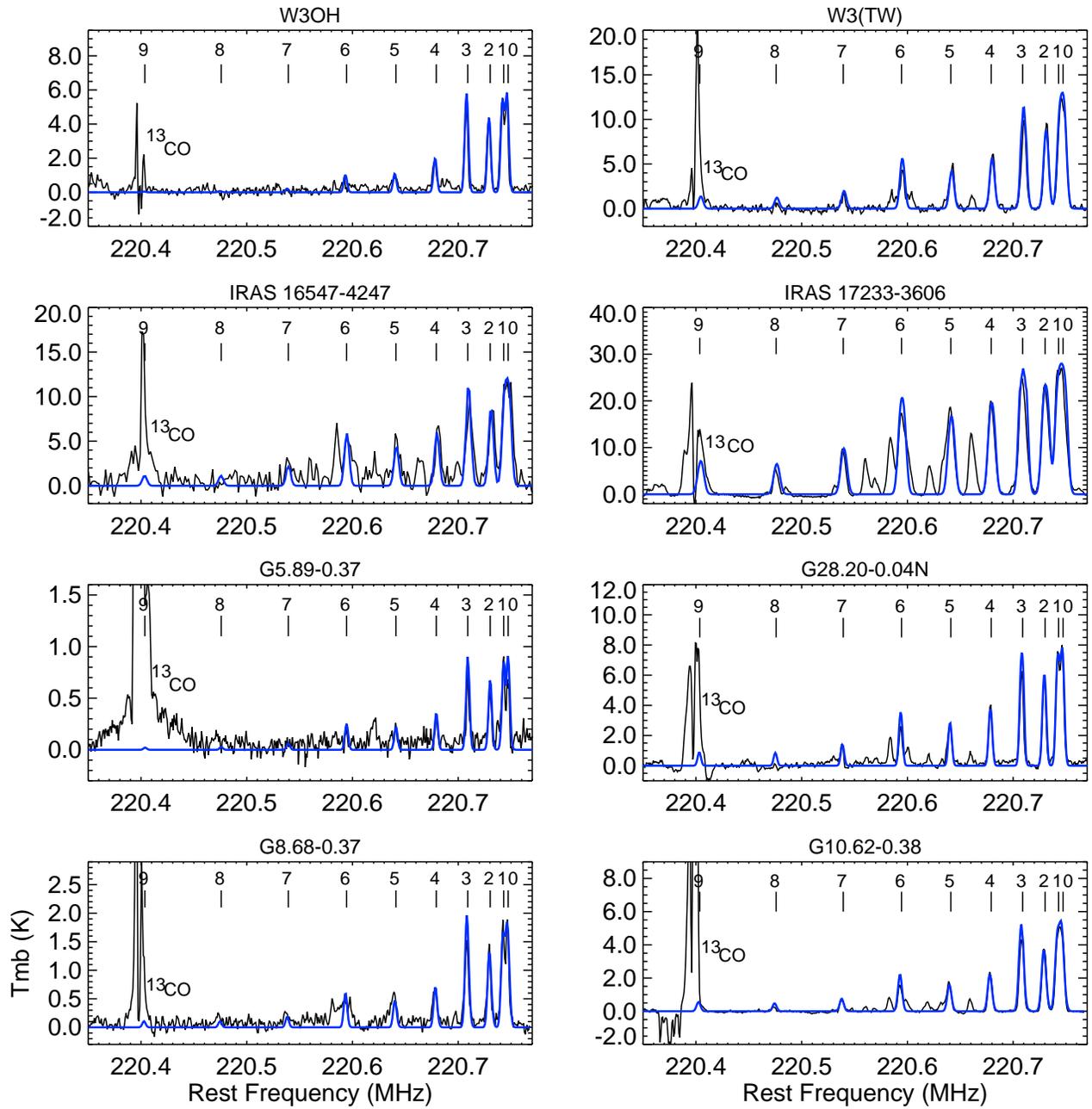}
\caption{Observed \chtcnte~spectra (black) and synthetic model (blue) spectra obtained with XCLASS. 
The fit parameters are given in Table 4. The numbers in each panel represent the $K-$ladder quantum numbers. 
The line at $\sim$220.4\,GHz is $^{13}$CO(2-1) and is overlapped with the $K=9$ line.\label{xclass1}}
 \end{figure*}

\newpage

 \begin{figure} 
 \centering
\includegraphics[angle=0,width=0.95\textwidth]{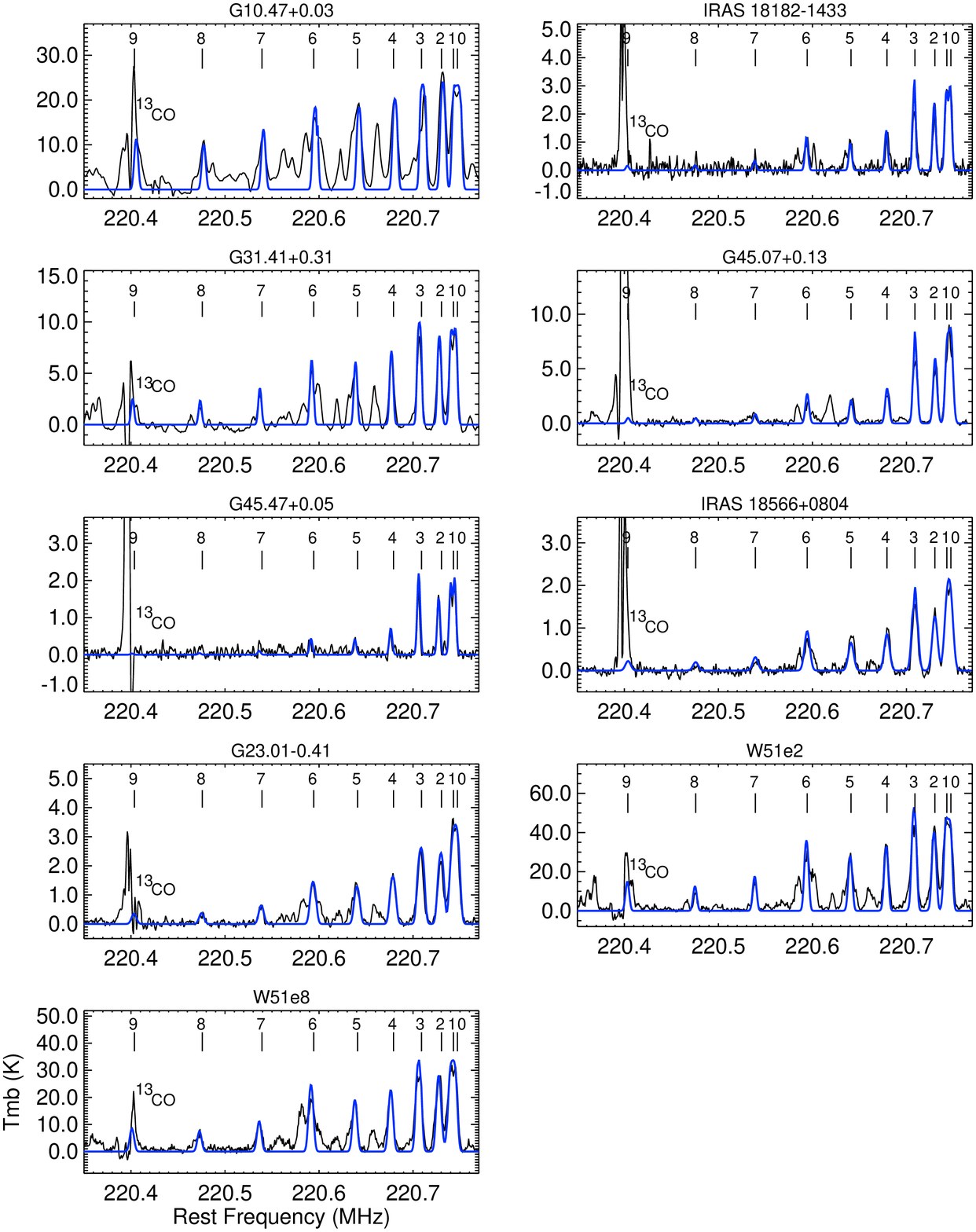}
  \caption{Continuation of Figure \ref{xclass1}.\label{xclass2}}
\end{figure}

\newpage

\begin{figure*} 
 \centering
\includegraphics[width=0.90\linewidth]{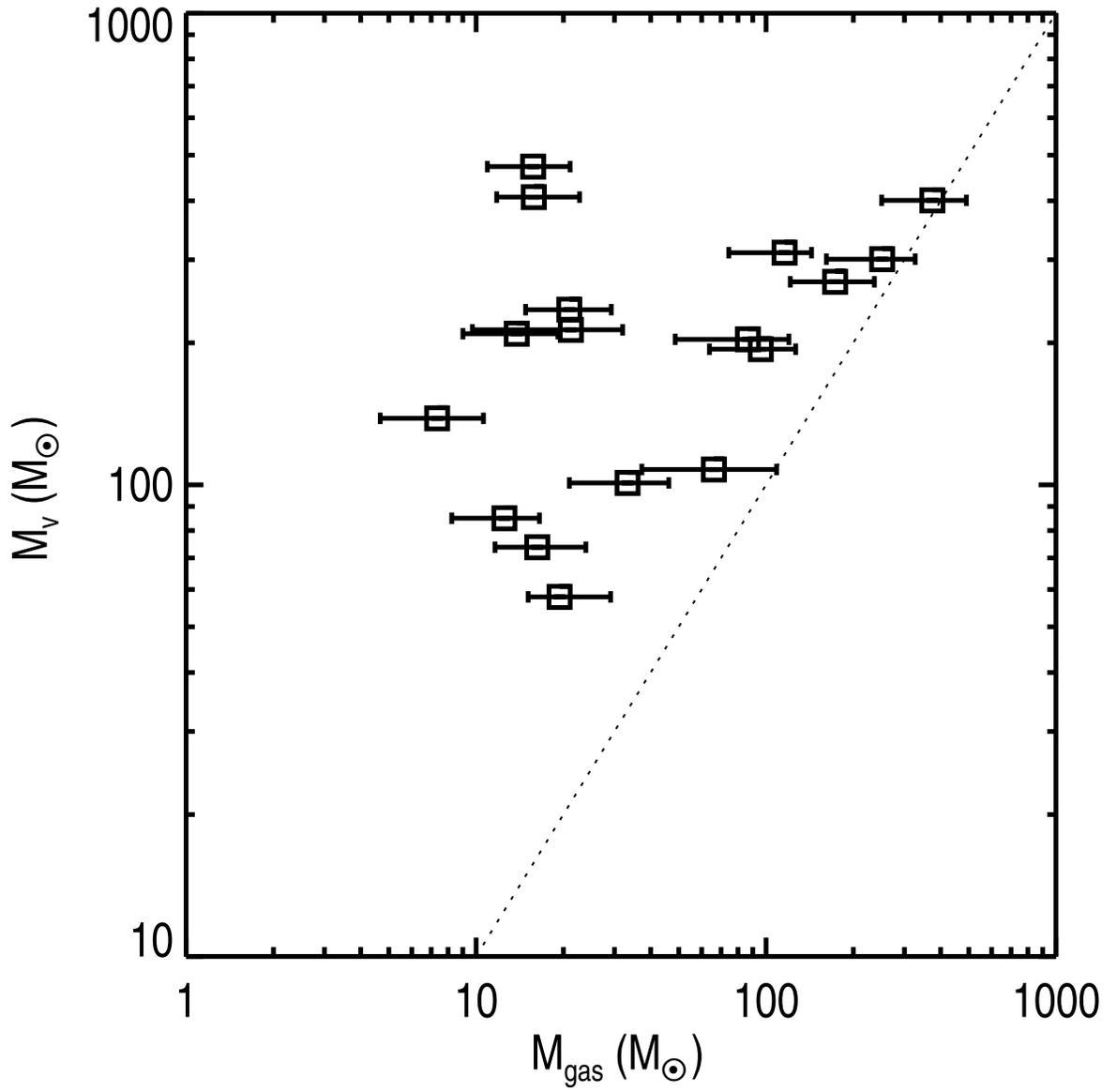}
\caption{Plot of virial mass versus gas mass. \mgas~ is estimated from the 1.3\,mm continuum emission 
           using the high temperature of the \chtcn~analysis (Table 5). $M_{\rm vir}$ is estimated using $\Delta V$ 
           from the observed linewidth (Table \ref{linedata}). The dotted line represents 
           $M_{\rm vir}$/\mgas~=1.\label{plots1}}
\end{figure*}
 
\newpage

 \begin{figure*} 
 \centering
 \includegraphics[width=0.90\linewidth]{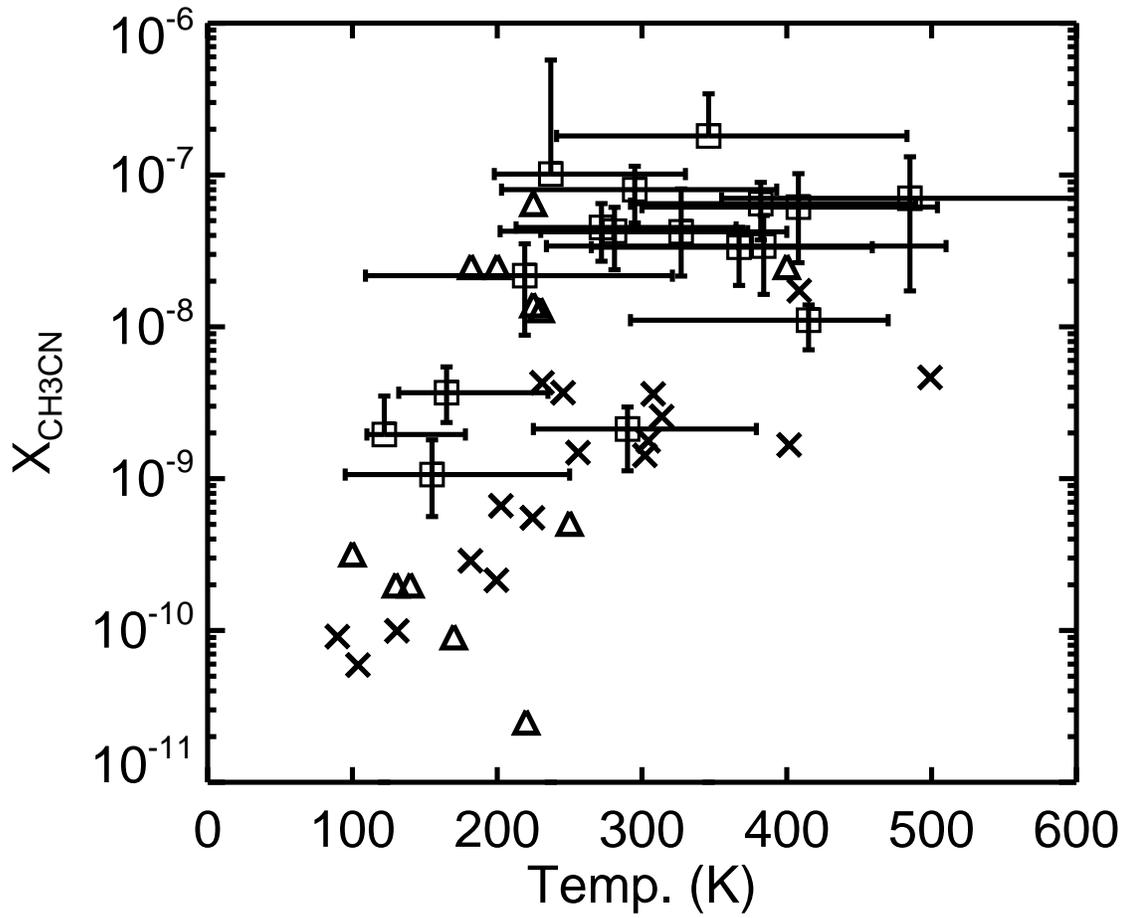}
  \caption{Fractional abundance of \chtcn~versus temperature of the compact hot and dense region estimated 
           with XCLASS (squares). Uncertainties of the hot components were used to estimate uncertainties 
           in gas masses and column densities from the continuum emission. For comparison, we plot our results 
           from the rotational diagrams (crosses) and reported values for other HMCs (triangles). Reported data 
           from the literature are taken from Galvan-Madrid et al. 2009, Chen et al. 2006, Wilner et al. 1994, 
           Zhang et al. 1998, and Wang et al. 2010.\label{plots2}}
\end{figure*}


\end{document}